\documentclass[fleqn,usenatbib]{mnras}

\usepackage[T1]{fontenc}
\usepackage{ae,aecompl}

\usepackage{graphicx}	
\usepackage{amsmath}	
\usepackage{amssymb}	
\usepackage{threeparttable}
\usepackage{multirow}

\newcommand{\ONE}{1E{\thinspace}1740.7$-$2942}
\newcommand{\GRS}{GRS{\thinspace}1758$-$258}
\newcommand{\ts}{\thinspace}
\newcommand{\gr}{$^{\circ}$}

\title[Background and Imaging Simulations for the Hard X-Ray Camera of the {\it MIRAX\/}
Mission]{Background and Imaging Simulations for the Hard X-Ray Camera of the {\it MIRAX\/} Mission}
\author[M. Castro et al.]{
M. Castro,$^{1}$\thanks{E-mail: manuel.castro@inpe.br}
 J. Braga,$^{1}$
 A. Penacchioni,$^{1}$
 F. D'Amico,$^{1}$
 and R. Sacahui$^{1,2}$
\\
$^{1}$Instituto Nacional de Pesquisas Espaciais, Av. dos Astronautas 1758, CEP: 12227-010, S\~ao Jos\'e dos Campos, SP, Brasil\\
$^{2}$Escuela de Ciencias F\'\i sicas y Matem\'aticas, Universidad de San Carlos de Guatemala, Ciudad Universitaria, zona 12, Guatemala\\
}

\date{Accepted XXX. Received YYY; in original form ZZZ}

\pubyear{2015}

\graphicspath{{./figs/}}

\begin{document}

\label{firstpage}
\pagerange{\pageref{firstpage}--\pageref{lastpage}}
\maketitle

\begin{abstract}

We report the results of detailed Monte Carlo simulations of the performance expected both at balloon altitudes and 
at the probable satellite orbit of a hard X-ray coded-aperture camera being developed for the {\it MIRAX\/} mission. 
Based on a thorough mass model of the instrument and detailed specifications of the spectra and angular dependence of 
the various relevant radiation fields at both the stratospheric and orbital environments, we have used the well-known 
package GEANT4 to simulate the instrumental background of the camera. We also show simulated images of source fields to 
be observed and calculated the detailed sensitivity of the instrument in both situations. The results reported here are 
especially important to researchers in this field considering that we provide important information, not easily found 
in the literature, on how to prepare input files and calculate crucial instrumental parameters to perform GEANT4 
simulations for high-energy astrophysics space experiments.
\end{abstract}




\section{Introduction}

Estimation of the energy spectrum of the background signal and its spatial distribution over the detector plane is 
crucial for the design of hard X-ray and gamma-ray astronomy telescopes. In the case of an observation of a point 
source from an orbital space platform (i.e. a satellite), the background consists in the diffuse electromagnetic 
radiation coming through the telescope aperture, emission from other sources in the field of view (FoV) and the 
instrumental background, which arises from interaction of high-energy particles with the detectors and surrounding 
material. In the case of observations carried out at stratospheric balloon altitudes, we also need to take into account 
the atmospheric radiation produced by the interaction of cosmic particles with atmospheric constituents, which will 
generate secondary particles and photons. In any case, the observations of cosmic X- and gamma-ray sources are always 
hindered by intense and complex background radiation measured by the detector system. In this paper we consider the 
contribution of photons, protons, electrons and neutrons to the total background of a hard X-ray coded-aperture imaging 
camera to be mounted first in a balloon platform and later in a satellite bus at near-equatorial Low-Earth Orbit (LEO).

In order to estimate the background, it is important to have an accurate knowledge of its origin. To achieve this, one 
needs to include detailed descriptions of the energy spectra and angular distribution of the particle fields that 
surround the instrument. In this work we describe the procedures we used in order to model the background of a hard 
X-ray imaging camera being developed at the National Institute for Space Research (INPE), Brazil, in the context of the 
{\it MIRAX} (Monitor e Imageador de RAios X) space astronomy mission \citep{2004AdSpR..34.2657B, 2006AIPC..840....3B}.

Since the instrument is supposed to fly on-board stratospheric balloons over Brazil, at an altitude of $\sim$ 42\ts km 
and a latitude of $\sim -23^{\circ}$ S, and as part of a satellite experiment in a near-equatorial circular LEO, we 
have included both environments in our simulations. We have developed a detailed mass model of the camera and simulated 
all the interactions in the instrument components and materials using the GEANT4 (GEometry ANd Tracking) code 
developed at CERN \citep{Agostinelli2003}. The results of these simulations have allowed us to estimate with good 
precision the spectral response of the CdZnTe (CZT) detectors of the camera and the spatial distribution of counts over 
the detector plane. Also, by running simulations with different configurations and analyzing the results, we were able 
to improve the passive shielding configuration of the camera in order to minimize the leakage of fluorescence radiation 
in the shield and achieve a near uniform distribution of the background over the detector plane.  This is extremely 
important for the quality of the reconstructed coded mask images. 

The balloon version of the instrument is called {\it protoMIRAX\/} \citep{Braga2015} and will be launched for the first 
time from Cachoeira Paulista, SP, Brazil, in late 2016 for a $\sim$ 24-hour flight. The balloon flight will test 
several {\it MIRAX\/} subsystems in a near-space environment and will carry out imaging demonstrations of the camera by 
observing bright X-ray sources.

\hspace{\textwidth}

This paper is organized as follows: In section \ref{sec:the X-ray camera} we describe the X-ray camera and present the 
mass model that we have developed to represent it in the simulations. In section \ref{sec:Input environments and 
fields} we describe the different particle and photon spectra, as well as their angular dependence, that we have used 
as inputs to the simulations of the various components of the background in both the high atmospheric and space 
environments. In section \ref{sec: Simulated background spectra and spatial distribution} we show the results of the  
simulations in terms of detector spectral response and the spatial distribution of counts over the detector plane. In 
section \ref{sec:imaging reconstruction} we present the results of imaging reconstruction using simulated observations  
of the Crab nebula and the Galactic Centre (GC) regions. Finally, in section \ref{sec:conclusion} we discuss the 
results obtained and present our conclusions. 

\section{The X-ray camera}\label{sec:the X-ray camera}

\subsection{Description}

The {\it protoMIRAX\/} balloon experiment consists of a wide-field coded-mask hard X-ray camera operating in a 
stratospheric balloon platform (gondola), together with auxiliary instrumentation. The experiment is described in 
detail by \cite{Braga2015}. For {\it MIRAX\/}, it is envisaged that a space-qualified version of the same camera will 
operate in a satellite platform. Even though {\it MIRAX\/} is an approved concept instrument for the Brazilian space 
program, the satellite bus that will carry {\it MIRAX\/} instruments are not yet fully defined.

The X-ray camera consists essentially of a coded mask, a passive shielding system surrounding the detectors, a 
collimator and a detector plane. The coded-mask is based on a Modified Uniformly Redundant Array (MURA) 13$\times$13 
basic pattern \citep{1989ApOpt..28.4344G}, extended to 25$\times$25 elements, placed 650\,mm away from the detection 
plane and supported by an acrylic frame. The mask elements are made of 1.0mm-thick lead, 1.5mm-thick tin and 
0.5mm-thick copper, each element having an area of 20$\times$20\ts mm$^2$. This will provide a $\sim 
20^{\circ}\times 20^{\circ}$ fully-coded field of view (FoV), with an angular resolution of $\sim 1^{\circ} 45'$. 
The detection plane consists of an array of 13$\times$13 CZT (Cadmium Zinc Telluride) crystals of dimensions 
10$\times$10$\times$2\ts mm with a gap of 10\,mm between two consecutive crystals. The energy range is from 10 to 
200 keV, although for the balloon version we will start from 30 keV due to the high atmospheric absorption below these 
energies. The full detection area is thus 169 cm$^{2}$. The collimator is a grid of crossed walls, 81\,mm high, made of 
lead, tin and copper layers in a configuration Cu(0.2\,mm)-Sn(1.0\,mm)-Pb(0.5\,mm)-Sn(1.0\,mm)-Cu(0.2\,mm). There is one 
detector at the bottom of each collimator cell, so that the centres of contiguous collimator blades are 20\ts mm apart. 
The collimator was designed to minimize background and define the total  FoV to match exactly the fully coded FoV defined by the mask. This concept has the advantage of having zero partially coded FoV, but, on the other hand, the instrument sensitivity falls off with the collimator response towards the edges of the FoV. With these parameters, as explained in \cite{Braga2015}, the total, fully-coded FoV of $20^{\circ}\times20^{\circ}$ has a $7^{\circ}\times 7^{\circ}$ region of full sensitivity and $14^{\circ}\times14^{\circ}$ full width at half maximum (FWHM).

\subsection{The mass model}

\begin{figure}
 \centering
\includegraphics[width=0.9\hsize]{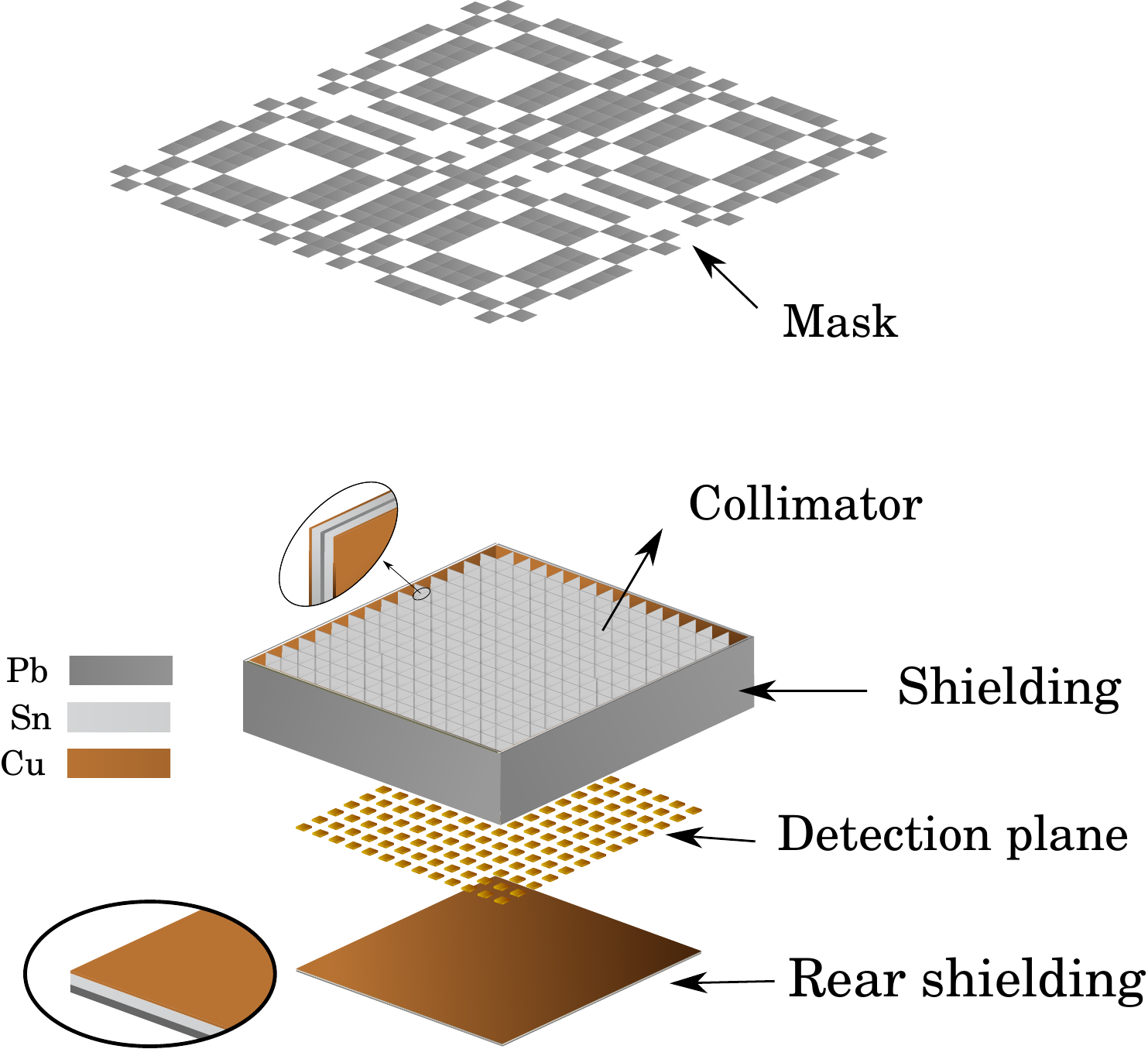}
\caption{Mass model of the X-ray camera as mounted in GEANT4.}
\label{MIRAX_mass_model}
\end{figure}

In order to run the Monte Carlo simulations using GEANT4, we need to have photons and particles coming from random 
directions and energies interact with the instrument, according to pre-defined spectral and angular distributions. 
Therefore, we need to build a detailed description of the overall geometry of the experiment and the material 
constituents of each part. We also have to define the geometric shape of each single piece, their dimensions, the 
materials they are made of and the relative position and orientation with respect to the other parts.  The code then 
simulates all the interactions of the primary and secondary particles with the different parts of the experiment before 
reaching the detectors. At the end of a simulation run, we have information on how many particles of each kind reach 
each of the detectors, and on the energy deposited by each particle. With this information we can build the event 
distribution over the detector plane for each energy range of interest. By considering the most important radiation 
fields present in the environments at which the experiment will be operating, we can calculate the contribution of each 
component to the total background to be measured by the experiment. Furthermore, we can include hard X-ray photons from 
cosmic sources of interest to build simulated shadowgrams of selected source fields and then produce simulated images 
of astrophysical regions. In Figure \ref{MIRAX_mass_model} we show the mass model of the hard X-ray camera as 
implemented in GEANT4.

\section{Environments and fields} \label{sec:Input environments and fields}

\subsection{The balloon environment}
\label{BkgprotoMIRAX}

At the stratospheric environment at which the balloon flights will be carried out (altitude $\sim 42$ km, latitude 
$\sim -23^{\circ}$), we have included in the simulations the main radiation fields: atmospheric $\gamma$-ray photons, 
electrons, atmospheric neutrons, and primary and secondary protons. The atmospheric $\gamma$-ray emission contribution 
has a dependence on the zenith angle $z$, so we have simulated each component separately according to the different 
incident angular ranges with respect to the zenith \citep{Gehrels1985} (see Table \ref{gammaback_proto}):

\begin{table}
\caption{Atmospheric $\gamma$-ray spectrum at an optical depth of 3.5 g\,cm$^{-2}$. The power-law spectra
 are in units of photons cm$^{-2}$ s$^{-1}$ sr$^{-1}$ MeV$^{-1}$.}
\label{gammaback_proto}
\centering
 \begin{tabular}{c|c|c}
\hline 
\hline
  Spectrum 			&	 Energy range (MeV) 		&	 Angular range ($^\circ$) \\
  \hline \hline
$2.19\times10^{2}\,E^{0.70}$	&	0.024 - 0.035			&	0 - 65 \\
$5.16\times10^{-2}\,E^{-1.81}$	&	0.035 - 10 			&	0 - 65 \\
$0.085\,E^{-1.66}$		&	0.1 - 10 			&	65 - 95 \\
$0.14\,E^{-1.50}$		&	0.1 - 10 			&	95 - 130 \\	
$0.047\,E^{-1.45}$		&	0.1 - 10 			&	130 - 180 \\
\hline 
\hline
 \end{tabular}
\end{table}
The electron spectrum in the energy range 1\,-\,10\,MeV at an atmospheric depth of 5 g\,cm$^{-2}$
can be modeled with a single power-law \citep{Gehrels1985}:
\begin{equation}
 \frac{dN}{dE}=1.4\times10^{-2}\,E^{-1.8} \,\,\,\,\,\,\,\,\, \rm{electrons\,\, cm}^{-2} \rm{s}^{-1} \rm{sr}^{-1} \rm{MeV}^{-1}
\label{electroneq}
\end{equation}

The proton contribution has two components, primary and secondary \citep{Dean2003}, depending on the energy range and 
whether the protons are part of the galactic cosmic rays or result from atmospheric interactions. 

The primary proton spectrum, in units of protons cm$^{-2}$ s$^{-1}$  MeV$^{-1}$, is well described by
\begin{equation}
 \frac{dN}{dE}=\begin{cases}
1.3\times10^{3}\,E^{-2}			& \text{ for }\,10000\,\text{MeV} \leq E \leq 30\,\text{GeV}	 \\
1.8\times10^{6}\,E^{-2.7}		& \text{ for }\,30\,\text{GeV} \leq E \leq 150\,\text{GeV}	
\end{cases}
\label{primaryprotoneq}
\end{equation}
whereas for the secondary protons the spectrum is given by
\begin{equation}
  \frac{dN}{dE}=\begin{cases}
     1.475\times10^{-6}\,E^{2}			&	\text{ for } \,5\,\text{MeV} \leq E \leq 16\,\text{MeV}		\\
 3.78\times10^{-4}\,E^{0}			&	\text{ for } \,16\,\text{MeV} \leq E \leq 100\,\text{MeV}	\\
 6.43\times10^{-3}\,E^{-0.61}			&	\text{ for } \,100\,\text{MeV} \leq E \leq 300\,\text{MeV}	\\
  1.78\,E^{-1.6}				&	\text{ for } \,300\,\text{MeV} \leq E \leq 6.6\,\text{GeV}	\\
\end{cases}
 \label{secondaryprotoneq}
 \end{equation}

The neutron background can be represented as follows \citep{Gehrels1985}:
\begin{equation}
 \frac{dN}{dE}=\begin{cases}
0.26\,E^{-1.3}				& 	\text{ for } \,0.8 \leq E \leq 10\, \text{MeV}			\\
6\times10^{-2}\,E^{-0.65}		&	\text{ for } \,10 \leq E \leq 100\, \text{MeV}			\\
  \end{cases}
\label{neutroneq}
\end{equation}
where $dN/dE$ is in units of neutrons cm$^{-2}$ s$^{-1}$ MeV$^{-1}$.

\begin{figure*}
 \includegraphics[width=0.49\hsize]{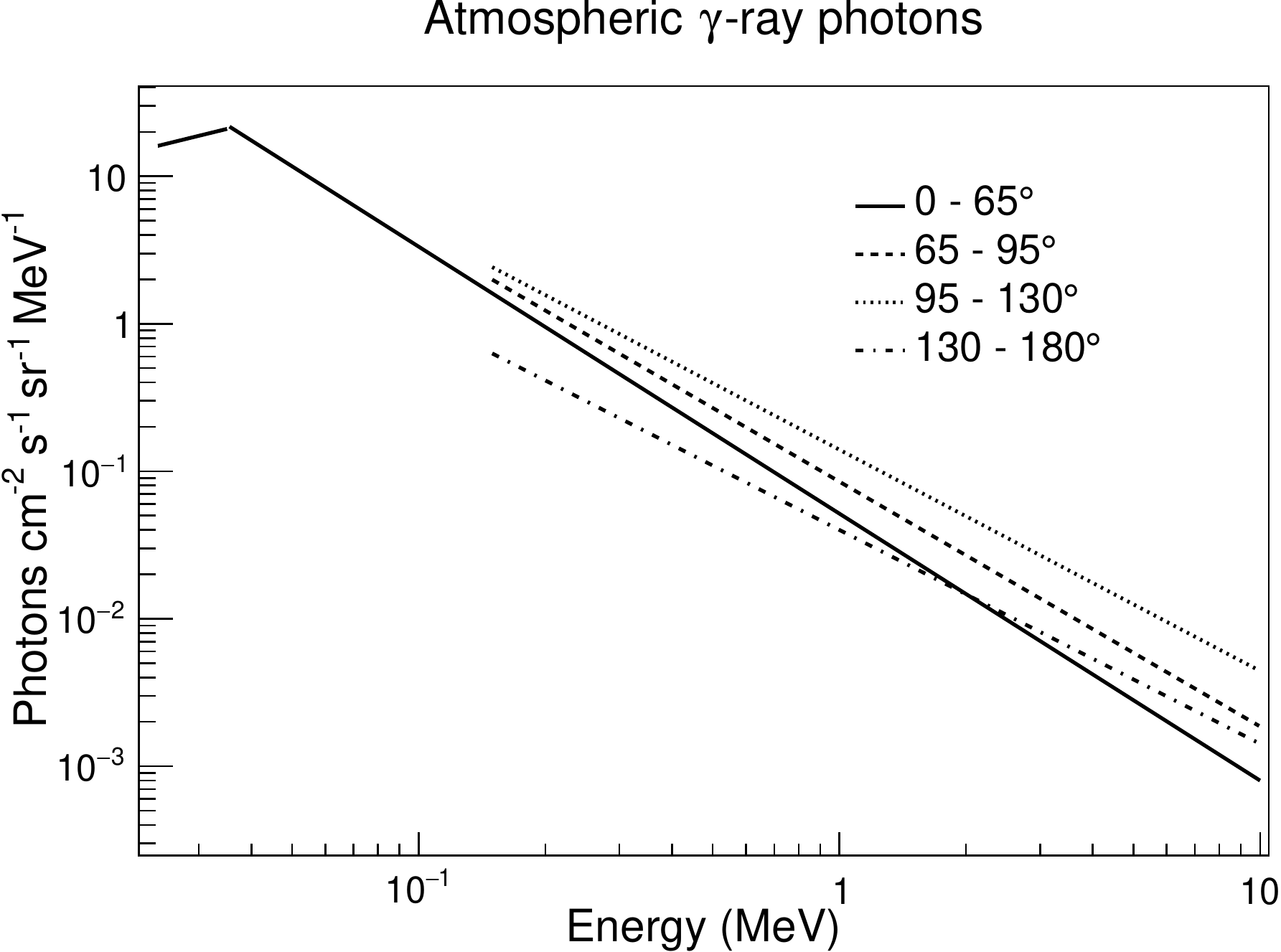} 
  \includegraphics[width=0.49\hsize]{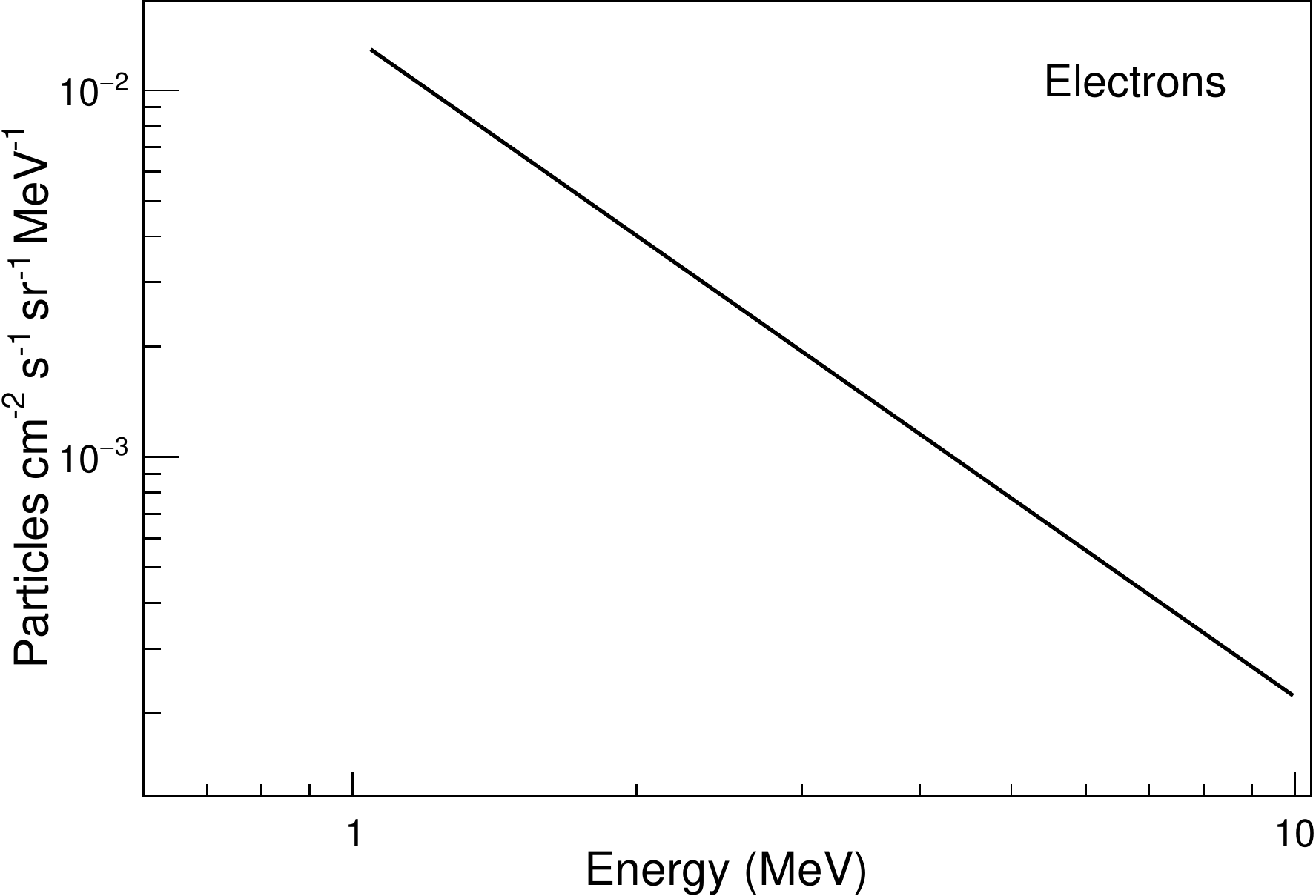} \\
\includegraphics[width=0.49\hsize]{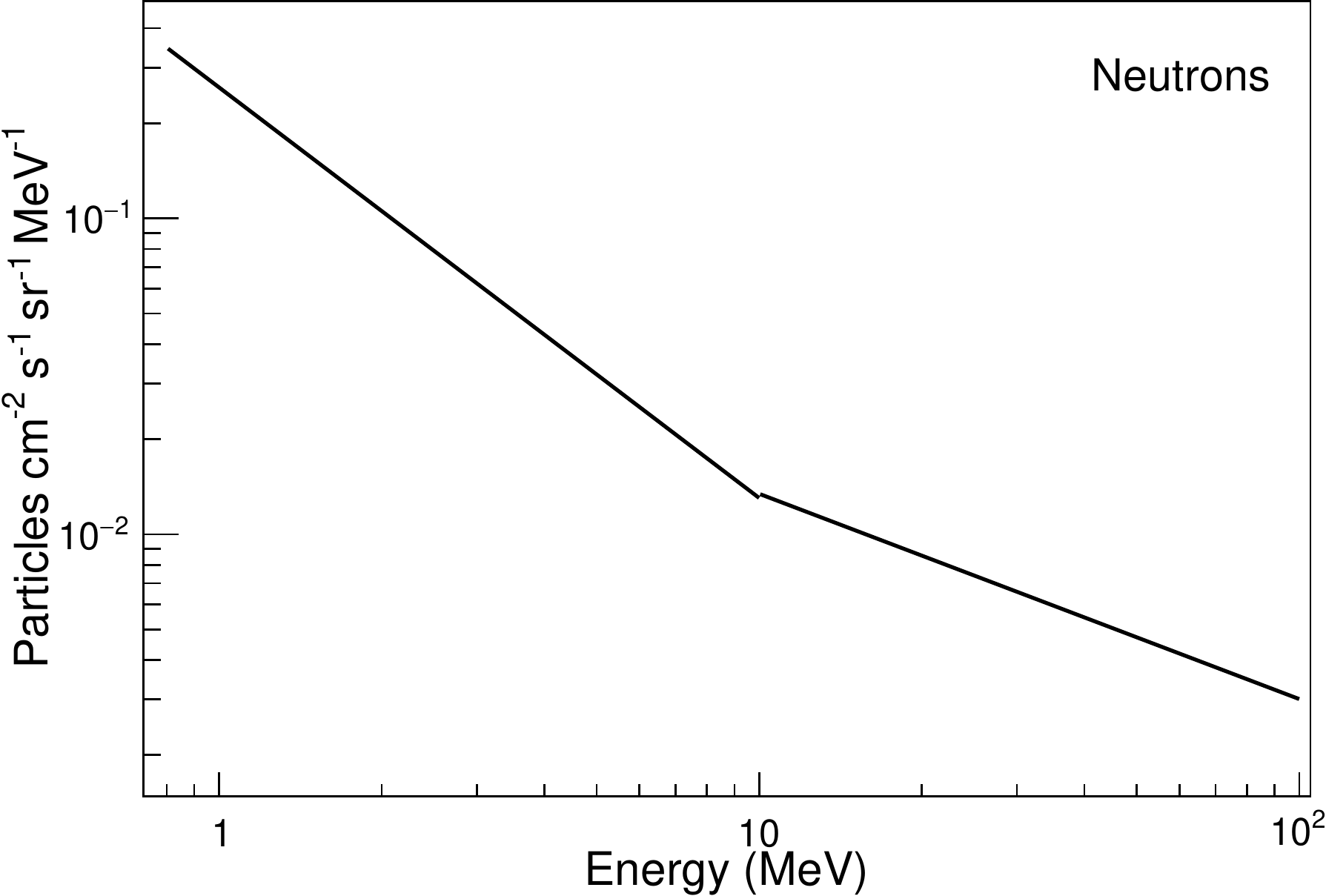}	  
 \includegraphics[width=0.49\hsize]{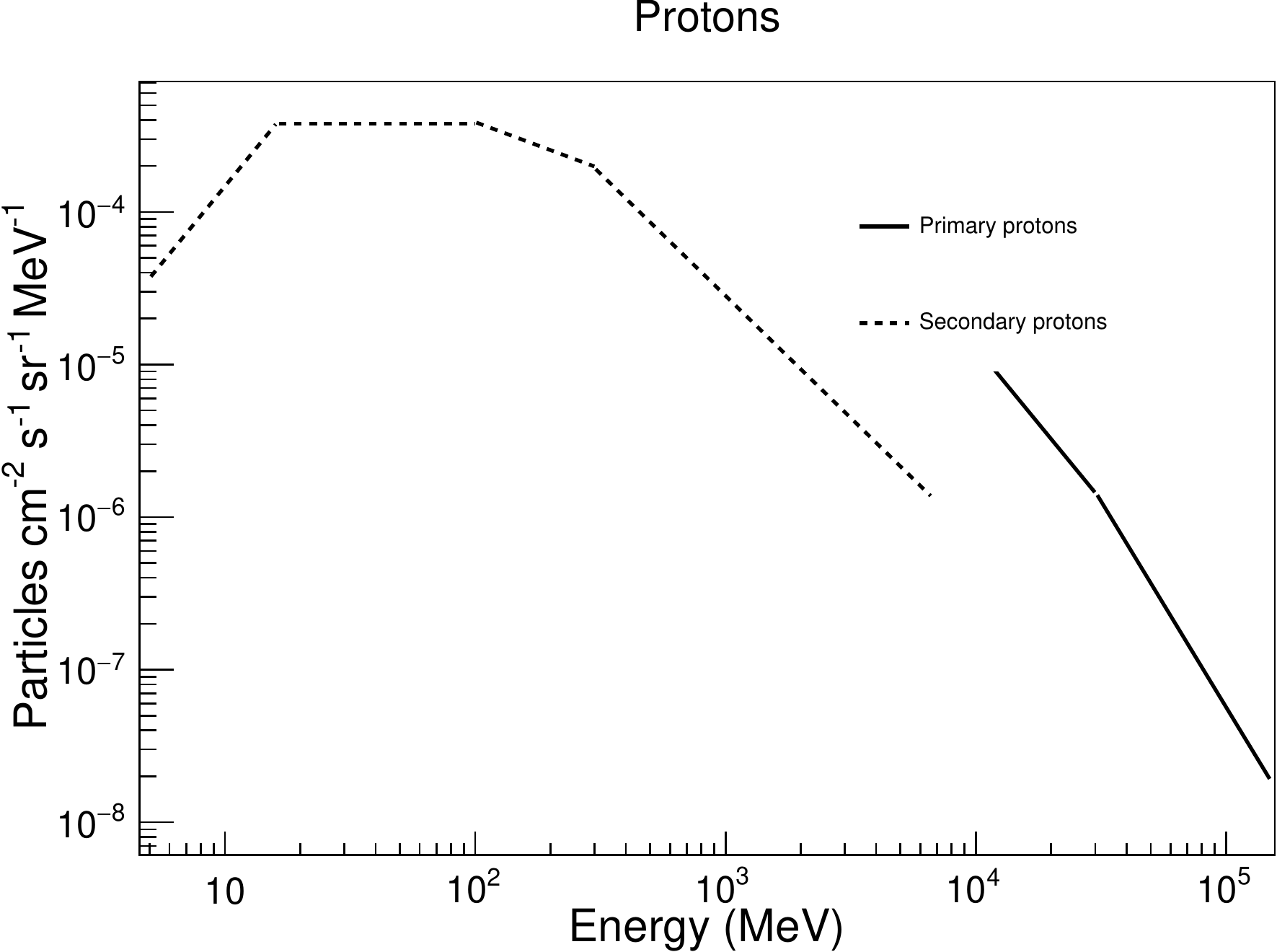} \\
\caption{Background sources at balloon altitudes. \textit{Top left}: $\gamma$-ray atmospheric emission. The solid line
corresponds to incident photons with zenith angles between 0$^{\circ}$ and 65$^\circ$; dashed to zenith angles between 65$^\circ$ and
95$^\circ$; dotted between 95$^\circ$ and  130$^\circ$, and dashed-dotted to the range 130$^\circ$ - 180$^\circ$. \textit{Top right}:
electron spectrum.  \textit{Bottom left}: neutron spectrum. \textit{Bottom right}: proton spectra; solid line refers to 
primary protons, dashed line to secondary protons.}
\label{fig:background}
\end{figure*}

Figure \ref{fig:background} shows the spectra for the different particle species that were used as input in the {\it 
protoMIRAX\/} background simulations.

\subsection{The {\it MIRAX\/} Low-Earth near-equatorial orbit environment}
\label{BkgMIRAX}
In order to simulate the background for the {\it MIRAX\/} observations, we have considered the main background sources 
at the expected satellite orbit. We exclude the satellite passages in the South Atlantic Anomaly (SAA), in which the 
particle fields are much more intense. We have considered the cosmic diffuse $\gamma$-ray radiation, albedo 
$\gamma$-ray photons, galactic cosmic rays (GCR), trapped protons and neutrons. The electron contribution is 
negligible.  
 
The cosmic diffuse $\gamma$-ray radiation spectrum between 10\,keV and 100\,MeV is given by \citep{Gruber1999}:
\begin{equation}
  \frac{dN}{dE}=\begin{cases}
7.877E^{-1.29}e^{-E/41.43}& \text{ for } \, E\leq60\,\text{keV} \\
     4.32\times10^{-4}\left(\frac{E}{60}\right)^{-6.5} +\\ 8,4\times10^{-3}\left(\frac{E}{60}\right)^{-2.58} +\\
     4.8\times10^{-4}\left(\frac{E}{60}\right)^{-2.05}&  \text{ for } \, E\geq60\,\text{keV} \\
\end{cases}
\label{gammaback}
\end{equation}

To simulate the albedo $\gamma$-ray photon field, we have used the spectrum, also between 10\,keV and 100\,MeV, given
by \cite{Ajello2008} and \cite{Sarkar2011}:
 \begin{equation}
 \frac{dN}{dE}=\begin{cases}
 \frac{1.87\times10^{-2}}{\left(\frac{E}{33.7}\right)^{-5}+\left(\frac{E}{33.7}\right)^{1.72}}& \text{ for } \,E\leq200\,\text{keV} \\ 
 1.01\times10^{-4}\left(\frac{E}{\text{MeV}}\right)^{-1.34} &\text{ for } \,200\,\text{keV} \leq E \leq20 \,\text{MeV}\\ 
 7.29\times10^{-4}\left(\frac{E}{\text{MeV}}\right)^{-2} & \text{ for } \,E \geq20\,\text{MeV}\\
\end{cases}
\label{albedo}
\end{equation}

The galactic cosmic ray spectrum in the energy range 10\,MeV - 20\,GeV is well described by \cite{Mizuno2004} and
\cite{Sarkar2011}:
\begin{align}
 \frac{dN}{dE}&= F\left(\frac{E}{GeV}\right)^{-a}\text{exp}\left(-\frac{E}{E1_{\rm cut}}\right)^{-a+1} + A\left(\frac{E+Ze\phi}{GeV}\right)^{-b} \times  \nonumber\\
&  \frac{(E+Mc^{2})^{2}-(Mc^{2})^{2}}{(E+Mc^{2}+Ze\phi)^{2}-(Mc^{2})^{2}}\times\frac{1}{1+\left(\frac{E}{E2_{\rm cut}}\right)^{-12}},
\label{GCR}
 \end{align}
where $e$ is the electron charge, $M$ is the particle's mass, $c$ is the speed of light, $Z$ is the atomic number,  $F$ = $1.23\times10^{-8}$ counts cm$^{-2}$ s$^{-1}$ sr$^{-1}$ keV$^{-1}$, $E1_{\rm cut}$=$5.1\times10^{6}$\,keV, $a$=0.155, A=$2.39\times10^{-6}$ counts cm$^{-2}$ s$^{-1}$ sr$^{-1}$ keV$^{-1}$, $b$=2.83, $\phi$=$6.5\times10^{5}$\,kV and $E2_{\rm cut}=1.13\times10^{7}$\,keV.

To simulate the trapped proton contribution from 100\,MeV - 6\,GeV we have used the following expression
\citep{Mizuno2004,Sarkar2011}
\begin{equation}
 \frac{dN}{dE}= F\left(\frac{E}{GeV}\right)^{-a}\text{exp}\left(-\frac{E}{E1_{\rm cut}}\right)^{-a+1},
\label{trappedprotons}
\end{equation}
with the constant values the same as above. 

Finally, the neutron spectrum between 10\,keV - 1\,GeV at these altitudes was modeled according to \citep{Armstrong1973}:
\begin{equation}
 \frac{dN}{dE}=\begin{cases}
9.98\times10^{-8}\left(\frac{E}{\text{GeV}}\right) ^{-0.5}& \text{ for } 10\,\text{keV}\leq E\leq1\,\text{MeV}\\
3.16\times10^{-9}\left(\frac{E}{\text{GeV}}\right) ^{-1.0} & \text{ for } 1\,\text{MeV}\leq E\leq100\,\text{MeV}\\
3.16\times10^{-10}\left(\frac{E}{\text{GeV}}\right) ^{-2.0} & \text{ for } 100\,\text{MeV}\leq E\leq100\,\text{GeV}\\
\end{cases}
\label{neutrons}
\end{equation}

For all the equations above (\ref{gammaback}, \ref{albedo}, \ref{GCR}, \ref{trappedprotons} and \ref{neutrons}),
$dN/dE$ is given in particles cm$^{-2}$ s$^{-1}$ sr$^{-1}$ keV$^{-1}$.

\begin{figure*}
\begin{tabular}{cc}
\includegraphics[width=0.49\hsize]{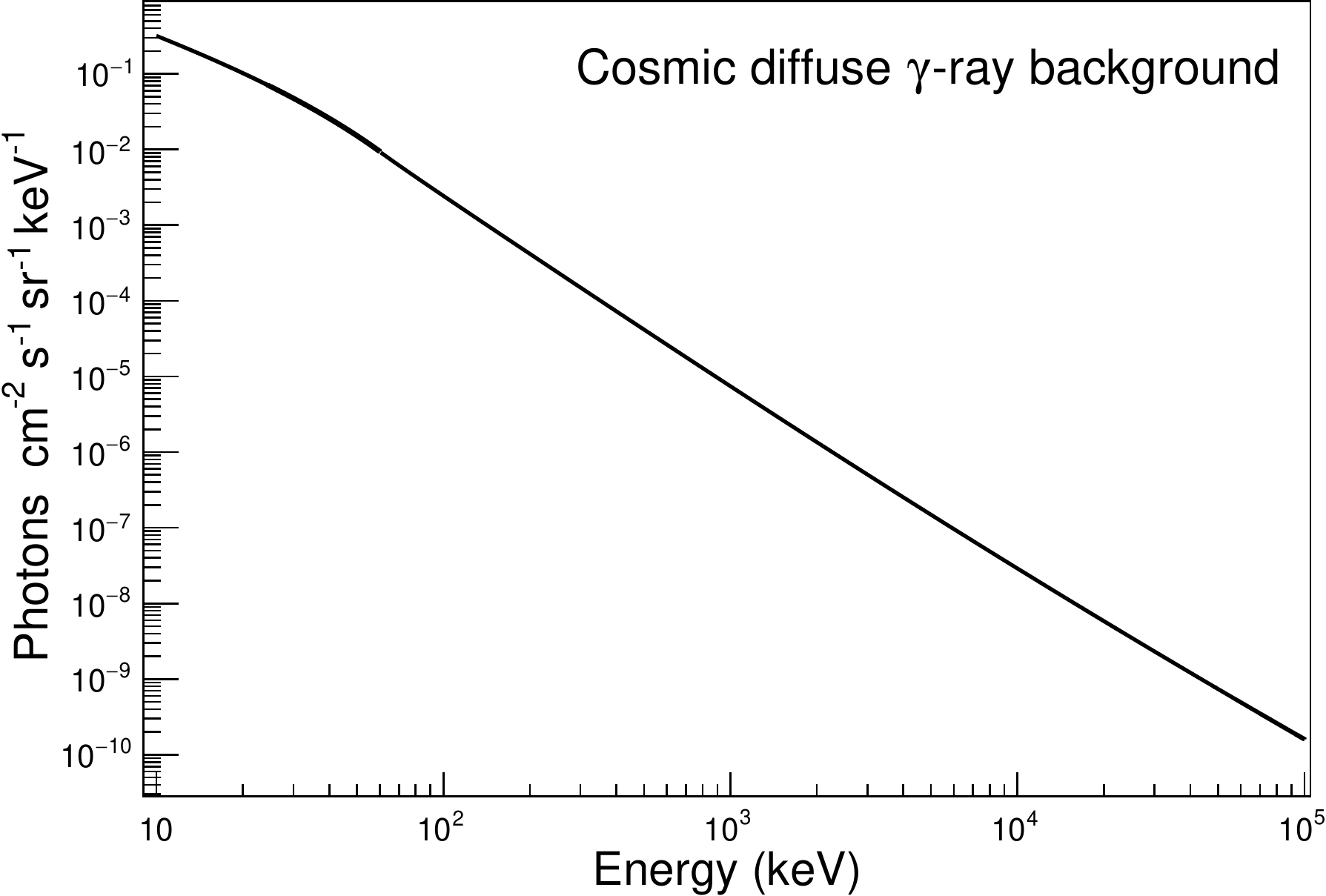}&\includegraphics[width=0.49\hsize]{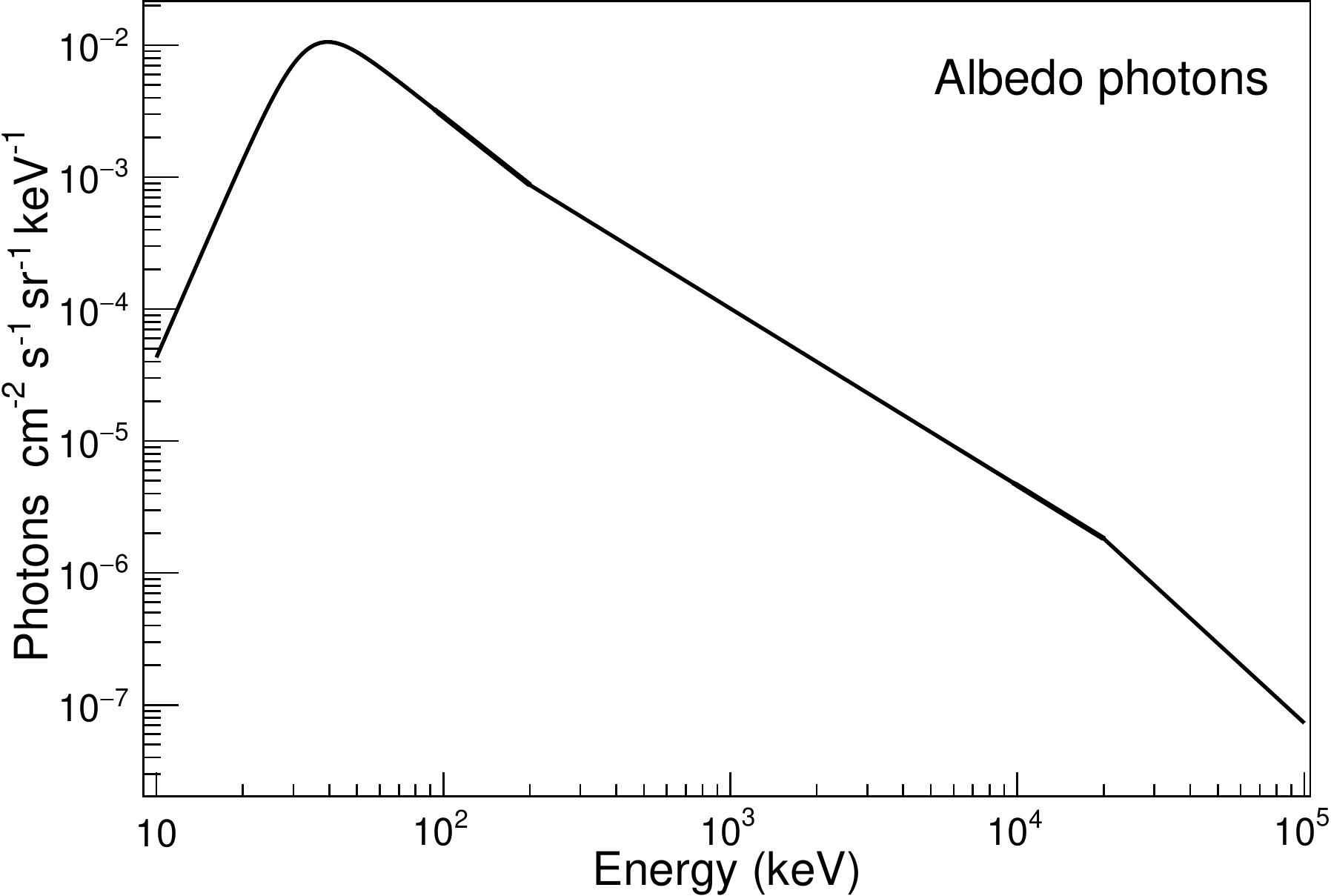} \\
\includegraphics[width=0.49\hsize]{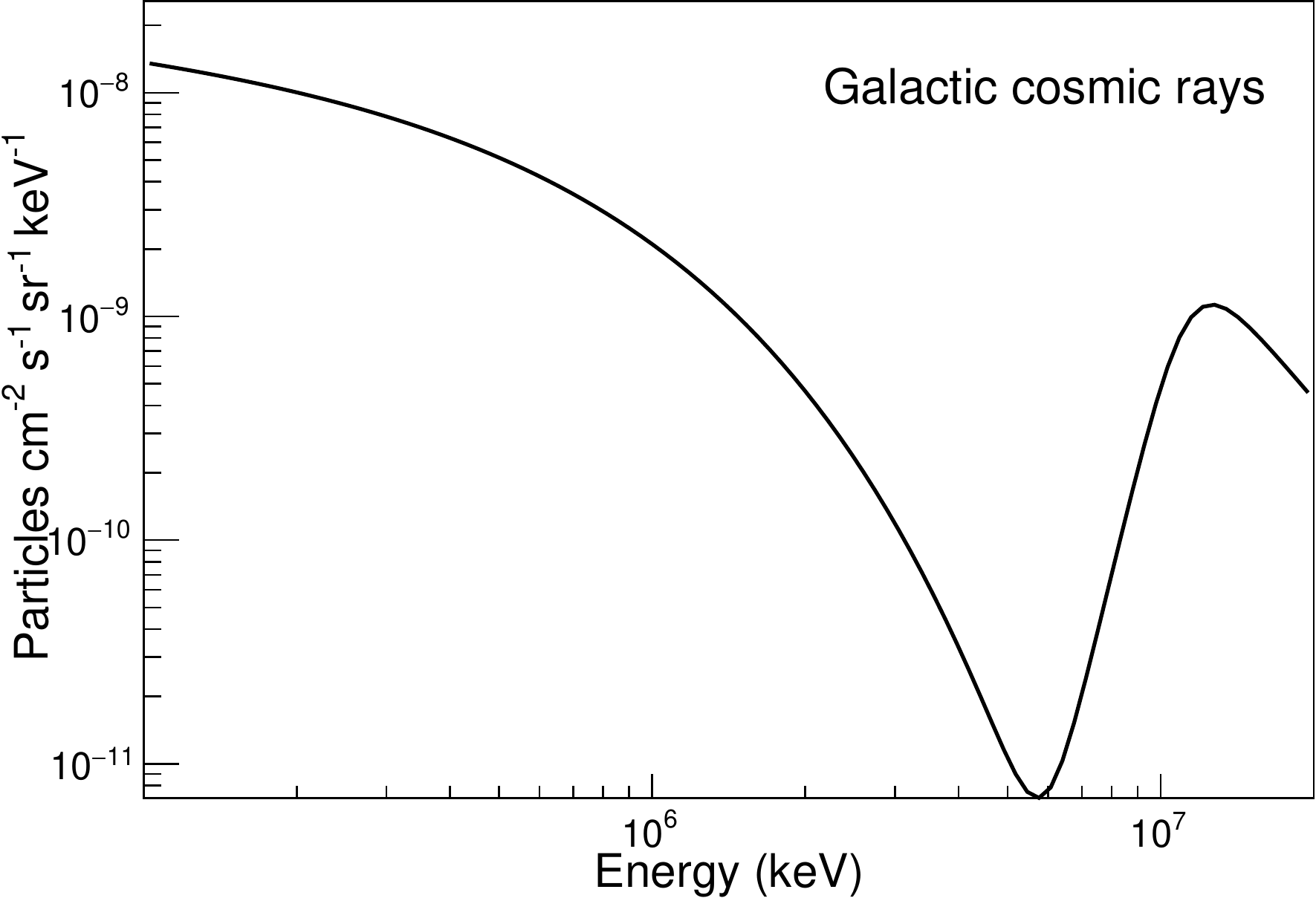} & \includegraphics[width=0.49\hsize]{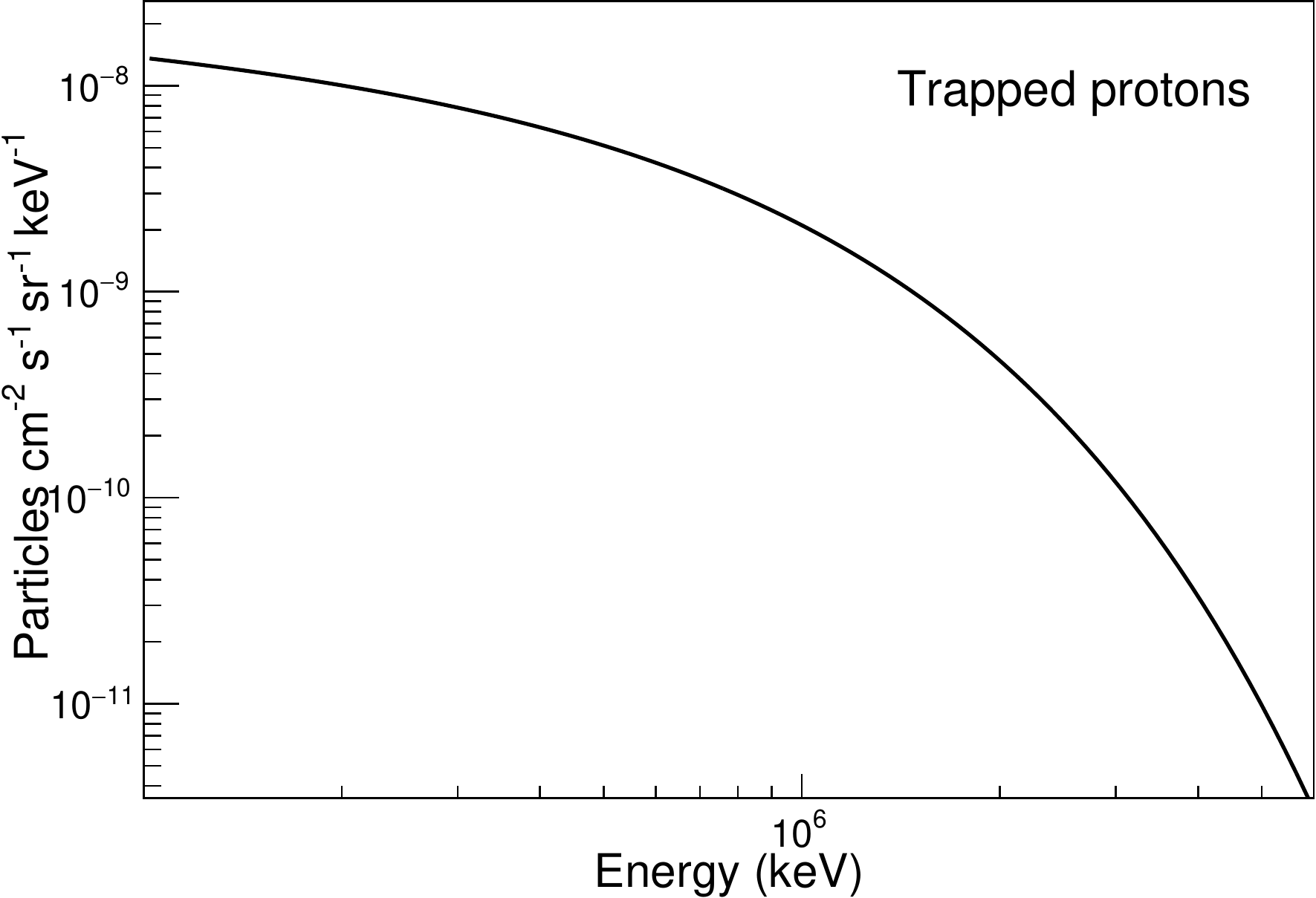} \\
\end{tabular}
\centering
\includegraphics[width=0.49\hsize]{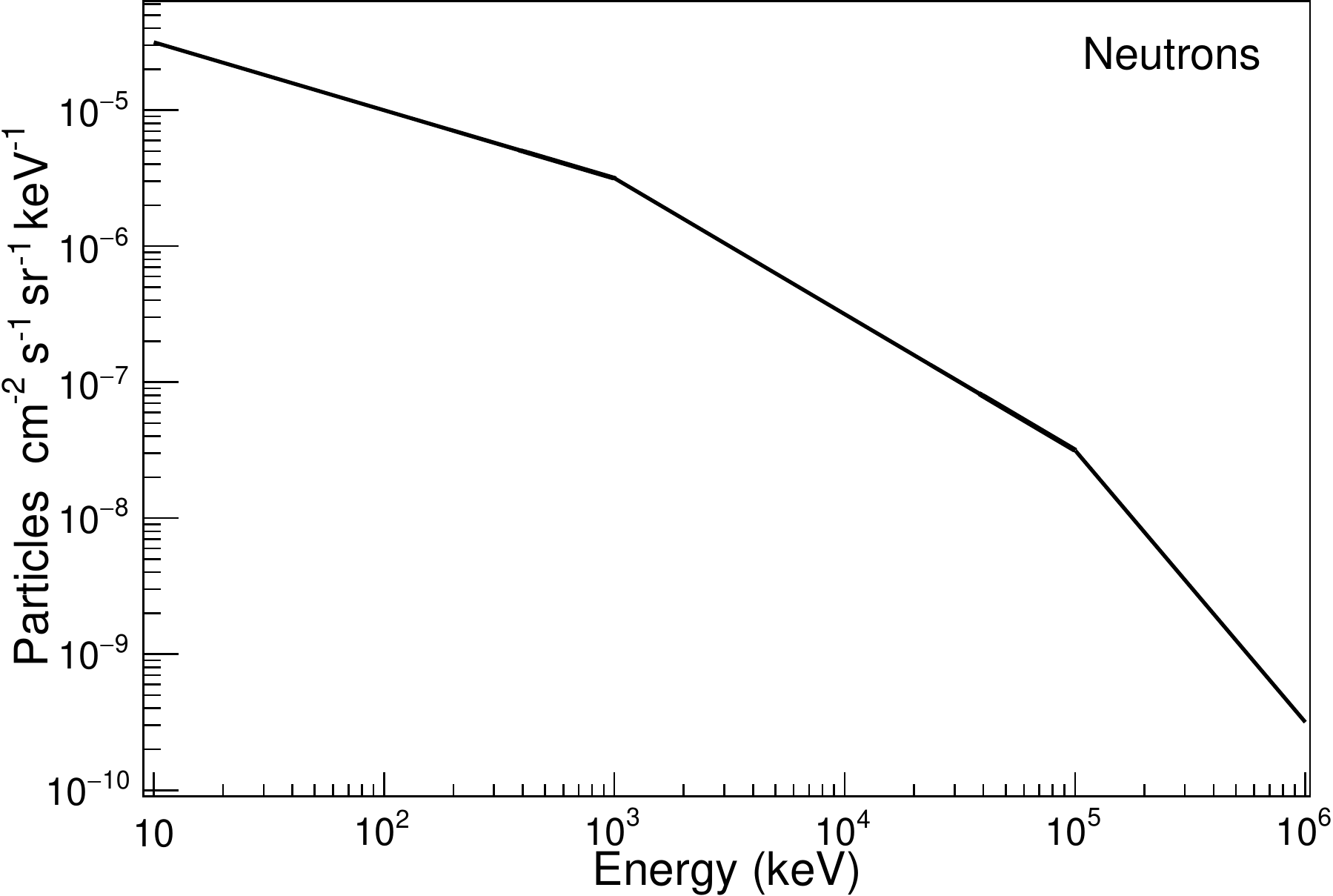}
\caption{Spectra of each radiation field present at the {\it MIRAX\/} orbit (outside the SAA), used as input in the
GEANT4 simulations of the background.}
\label{fig:input spectra MIRAX}
\end{figure*}

Figure \ref{fig:input spectra MIRAX} shows the spectra for the different kinds of particles in the respective energy
ranges.

\section{The simulated background} \label{sec: Simulated background spectra and spatial distribution}

Using the mass model and the radiation spectra described above as inputs to GEANT4, we performed simulations to study 
in detail the instrumental response of the X-ray camera under the two environments described in the previous section.

For each radiation field, we have used closely-spaced data points from the original spectrum to run separate 
simulations for each component. GEANT4 runs internal interpolation tasks to generate random events following the 
spectral behaviour of the original spectrum. In order to comply with the GEANT4 specific input procedures, we have 
built a model in which the X-ray camera is placed at the centre of a virtual emission sphere. The emissions from the 
spherical surface are then randomly created by the program both in terms of energy and directions, always pointing to 
the internal volume of the sphere. Primary particles coming from space will reach the experiment from any direction 
between $z=0$ and $z=\pi/2$, where $z$ is the zenith angle, while secondary particles, generated by the 
interaction of primary particles with the atmosphere, can reach the instruments from all directions. For {\it 
protoMIRAX\/}, the atmospheric gamma radiation spectra are generated in spherical sectors bounded by the incident angle 
ranges described above, while the primary protons come from the upper hemisphere and the neutrons, secondary protons 
and electrons come equally from all directions. In this case the radius of the virtual sphere (300\ts cm) was very large large compared to the instrument dimensions in order to insure that the randomly generated events in the internal surface of the sphere would follow closely the angular distribution of the radiation field without introducing significant effects due to the local geometry.
For the {\it MIRAX\/} environment, all the events hitting the detector are coming from a hemisphere, either centred on the zenith (diffuse $\gamma$-rays and cosmic rays) or the nadir (the other three fields). The radius of the sphere used in this case (90 cm) did not have to be too large since effects introduced by the local geometry are not significant.

From the GEANT4 output files we can determine the deposited energy in each of the detectors by each kind of interaction 
chain in the instrument. The output of the simulations allows us to build a spectrum and a spatial distribution of 
events over the detector plane for each of the radiation fields that interact with the instrument and for each of the 
169 CZT detectors. Those components can then be added up to represent the total background to be measured by {\it 
protoMIRAX\/} and {\it MIRAX\/}. This separation of the background contributions is extremely useful for the design of 
the structure and geometry of the instrument's passive shielding. By using the best shielding configuration, we can 
homogenise the spatial distribution of counts over the detector plane and lower the overall background. This is 
essential to improve the instrument sensitivity. 

\subsection{Normalisation}

In the simulations using GEANT4, the user specifies the number $N$ of virtual particles that will be created, either 
coming from random directions or from user-specified directions. In the case of a point source at infinity, like for 
instance a stellar X-ray source, the photons come in parallel lines at random positions inside a cylinder that 
encompasses the whole instrument. The lines are all parallel to the cylinder's axis. For radiation fields that have 
angle dependence, the user has to specify a spherical surface from which the particles will come, as explained in the previous section. GEANT4 creates virtual particles at random positions in the surface of the sphere and with random directions, provided they all point to the inside volume of the sphere. The program then creates the secondary, tertiary etc.\ particles that are produced by all interactions in the instrument mass model and tracks all their directions. Finally, it computes the energy losses on the detectors produced by the interactions of all particles that hit each specific detector's volume.

\begin{figure}
 \includegraphics[width=\hsize]{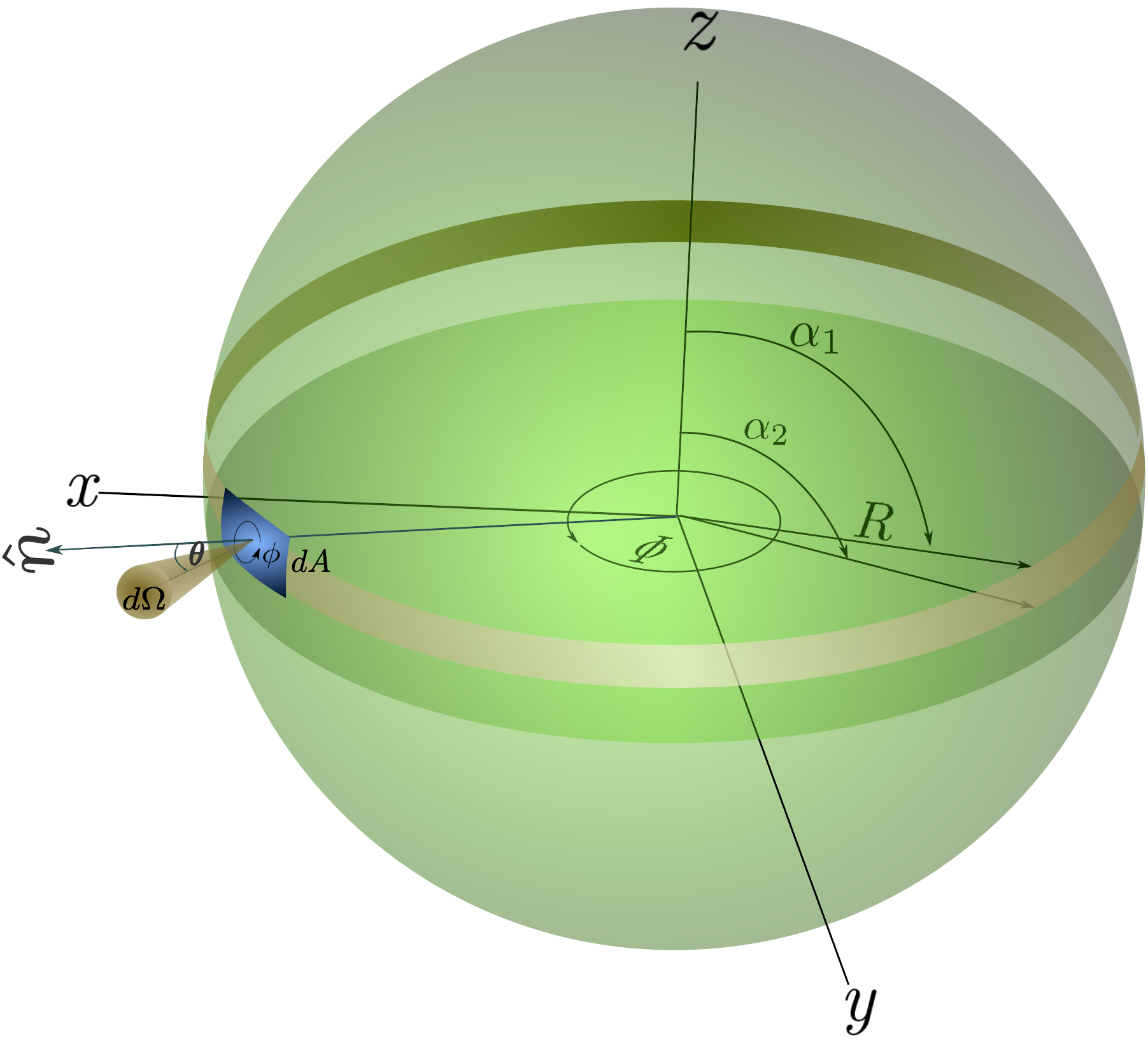}
\caption{Geometry used for the normalisation of GEANT4 simulations.}
\label{sphere}
\end{figure}

One important question that arises is how to relate the integration time of the simulated observation to the value of $N$. 
In our simulations, we have used the following procedure. The geometrical scheme is shown in Figure \ref{sphere}. We 
first note that the number of particles $dN$ passing through the area $dA$ within a solid angle $d\Omega$ during a time 
$dT$ in the energy range $dE$ is $dN=I_{N}\,dA\,dT\,d\Omega\,dE,$ where $I_{N}$ is the specific intensity of the 
radiation field in units of particles cm$^{-2}$ s$^{-1}$ sr$^{-1}$ keV$^{-1}$.  We now define an element of area $dA$ 
in an arbitrary direction $\hat{n}$. The net differential flux from the solid angle $d\Omega$ is given by 
$dF_{N}=I_{N}\cos{\theta}\,d\Omega$. The net flux is then obtained integrating $dF_{N}$ over all the solid angles,
\begin{equation}
 F_{N}=\int I_{N} \cos{\theta}\,d\Omega.
\end{equation}

Now, the total number of particles $N$ created in the sphere (or a spherical sector) to represent this flux will be
\begin{equation}
 N=\int I_{N}\,\cos{\theta}\,d\Omega\,dA\,dt\,dE,
\label{Nparticles}
\end{equation}
with $d\Omega=d\phi\,\sin{\theta}d\theta$, $dA=R^2\sin\alpha d\alpha d\Phi$ (see Figure \ref{sphere}), $0\,\leq\,\theta\,\leq\,\pi/2$ and $0\,\leq\,\phi\,\,\leq\,2\pi$. $R$ is the radius of the sphere and the range of $\alpha$ will depend on the specific spherical sector related to the incident angles of the radiation field.

We then have
\begin{equation}
 N=\pi\,A\,T\int I_{N}\,dE,
\label{events}
\end{equation}
where $A$ is the geometrical area of the spherical sector and $T$ is the total integration time. Finally, the integration
time is obtained by
\begin{equation}
T=\frac{N}{\pi\,A\,\int I_{N}\,dE},
\end{equation}
where $I_{N}=dN/dE$ is directly related to the radiation field spectra of sections \ref{BkgprotoMIRAX} and  \ref{BkgMIRAX}.
For the {\it protoMIRAX\/} experiment we have considered an 8-hour integration time in order to have statistics compatible
with what is achieved in timescales of balloon flights. For the satellite version, each radiation spectrum
(see Figure \ref{fig:background}) was considered separately so as to have good enough statistics for a good estimation
of all background components. 

\subsection{Background spectra}
\label{bkg_spectra}

\begin{figure}
\centering
\includegraphics[width=\hsize]{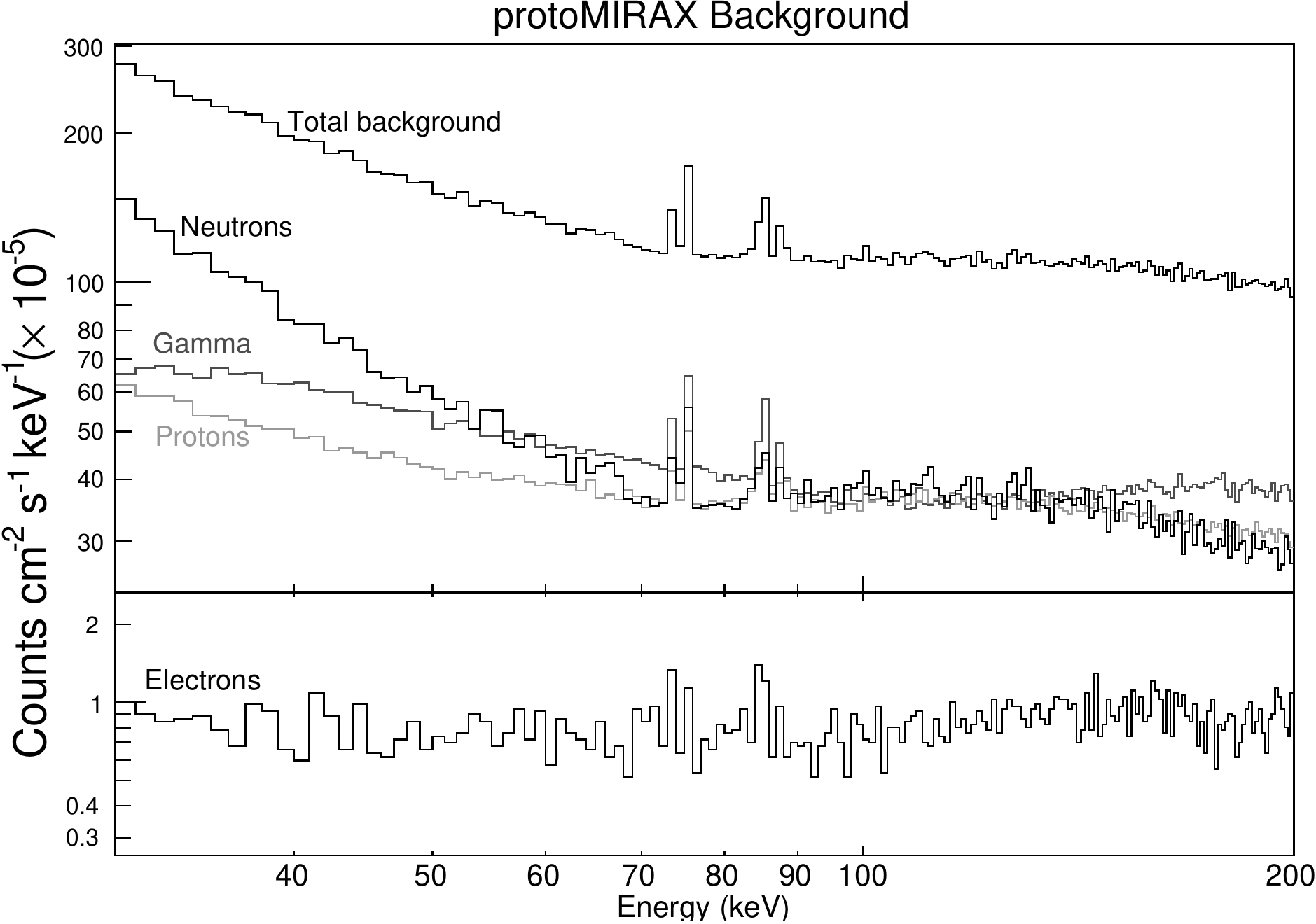}
\caption{Total background spectrum measured by {\it protoMIRAX\/}.}
 \label{fig:background spectrum protoMIRAX}
\end{figure}

In Figure \ref{fig:background spectrum protoMIRAX} we show the simulated spectra produced by the several radiation
components at balloon altitudes. One can see that the contribution of X-rays from atmospheric "fast" neutron interactions
in the instrument is very important below $\sim$50 keV, whereas the contributions of photons, protons and neutrons are
very similar between $\sim$100 and $\sim$150 keV. The atmospheric diffuse gamma radiation is the strongest component
above $\sim$150 keV. The contribution from electrons, about two orders of magnitude weaker, is clearly negligible. The
fluorescence lines of lead are clearly visible at 73, 74, 84 and 85\,keV\footnote{\url{http://www.kayelaby.npl.co.uk/atomic\_and\_nuclear\_physics/4\_2/4\_2\_1.html}}.
We have changed the thicknesses of the passive shielding materials (Pb-Sn-Cu) several times until we reached the best
solution for the instrument in terms of leakage of the fluorescence lines without adding too much weight.

\begin{figure}
 \includegraphics[width=\hsize]{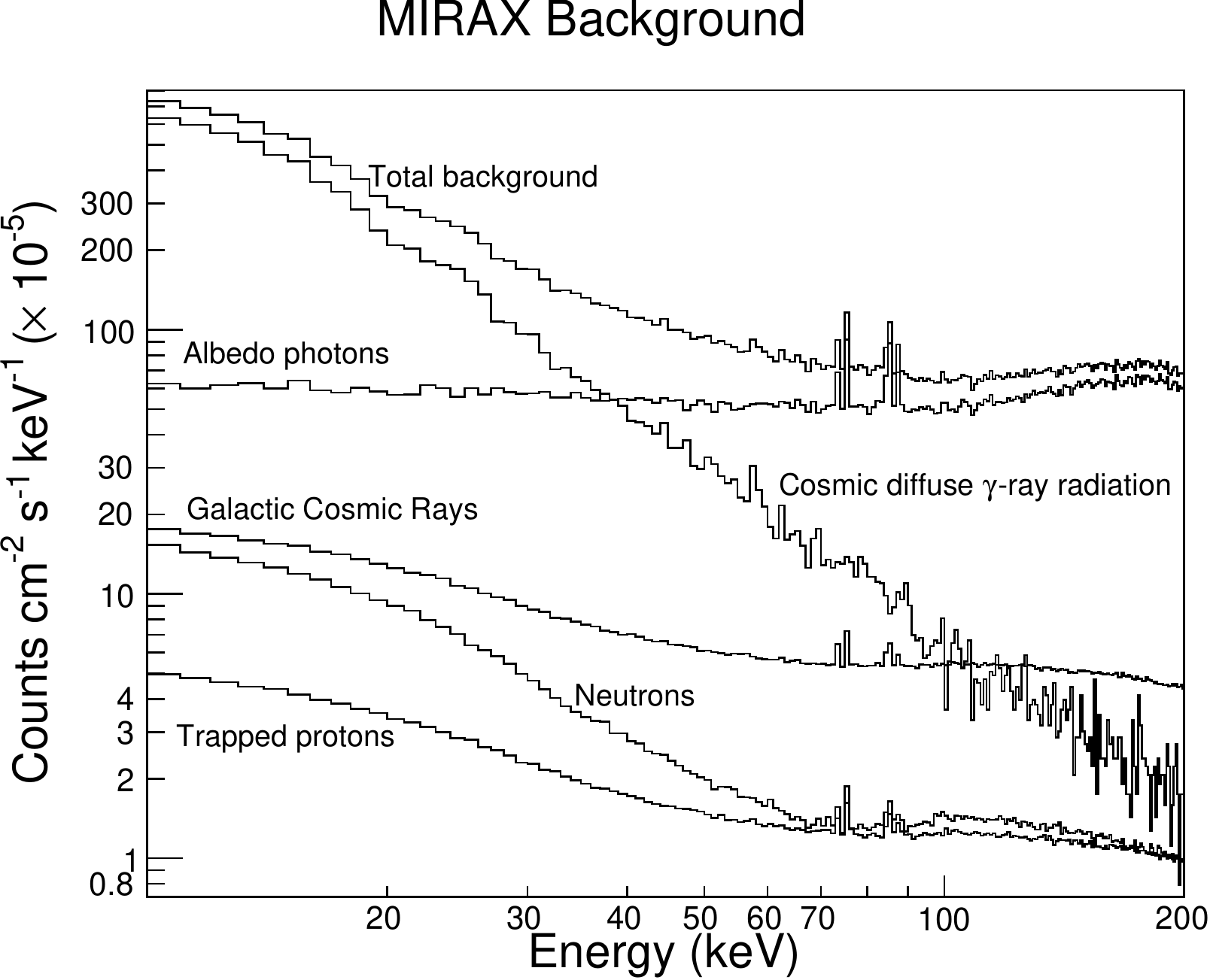}
\caption{Total background spectrum measured by {\it MIRAX\/}. The contribution of each individual component is also shown.}
\label{BackgroundMIRAX}
\end{figure}

In figure \ref{BackgroundMIRAX} we show the camera's background spectrum at the orbit of MIRAX. The cosmic diffuse 
radiation is the dominant component up to $\sim$40 keV. The interactions of albedo photons from the atmosphere below 
contribute very strongly and becomes a very important component all the way to 200 keV. Galactic cosmic rays 
(essentially protons) are also important, albeit about one order of magnitude weaker. It is important to mention that 
currently we are optimising the X-ray camera for protoMIRAX. After the balloon flights, we will make the necessary 
changes in the configuration so as to optimise the camera for operation in orbit. The contribution of albedo photons 
will certainly be greatly reduced by a different shielding system and by spacecraft materials as well. The simulations 
reported here will be of utmost importance to the final design of the camera to operate in space.

\subsection{Background spatial distribution over the detector plane}

A very important result of our simulations is the spatial distribution of background counts across the detector plane. 
In the case of {\it protoMIRAX\/}, we have taken into account the fact that the gamma-ray flux at the upper atmosphere is 
highly anisotropic. In order to obtain an accurate description of what happens at the detector level for each of the 
angular components, we produced count maps for each component (first four plots in Figure \ref{fig:shadowgram 
background protoMIRAX}). In the $0^{\circ}-65^{\circ}$ range, one can see the effects of the collimator response: since 
the flux from $0^{\circ}$ to $3^{\circ}$ is higher than at larger angles due to the collimator response (see 
\cite{Braga2015}), the shadow of the central bar on the mask pattern dominates so that we have fewer counts on the 
detectors underneath it. We have confirmed this hypothesis by rotating the mask 90\gr. For the $65^{\circ}-95^{\circ}$ 
range, we see a deep nonuniform distribution due to the fact that the detectors near the corners and edges of the 
detector plane are less protected by the collimator blades and shielding materials. $\gamma$-rays coming from below the 
instrument have a nearly homogeneous distribution, as expected. The distribution caused by primary protons shows less 
counts towards the edges and corners. This is also expected, since the probability for production of secondary 
radiation increases with the number of collimator blades the protons interact with. This is less pronounced with 
secondary protons, since they come from all sides. The atmospheric neutrons and electrons produce a more uniform 
distribution, within statistics. If we add all components of the background (Figure \ref{fig:total background 
shadowgram}), we get a very uniform distribution. This was also a consequence of the particular configuration we have 
implemented for the collimator and shielding materials, as reported by \cite{Braga2015}. The total (30--200\ts keV) 
background count rate for {\it protoMIRAX\/} is estimated to be 35.5\ts counts s$^{-1}$. It is noteworthy that 
preliminary results reported by \cite{Penacchioni2015} have provided a count rate of $\sim$50 counts s$^{-1}$. The 
improved mass model we built based on preliminary simulation results have led to this 29\% decrease in the background 
level.

\begin{figure*}
\includegraphics[width=\hsize]{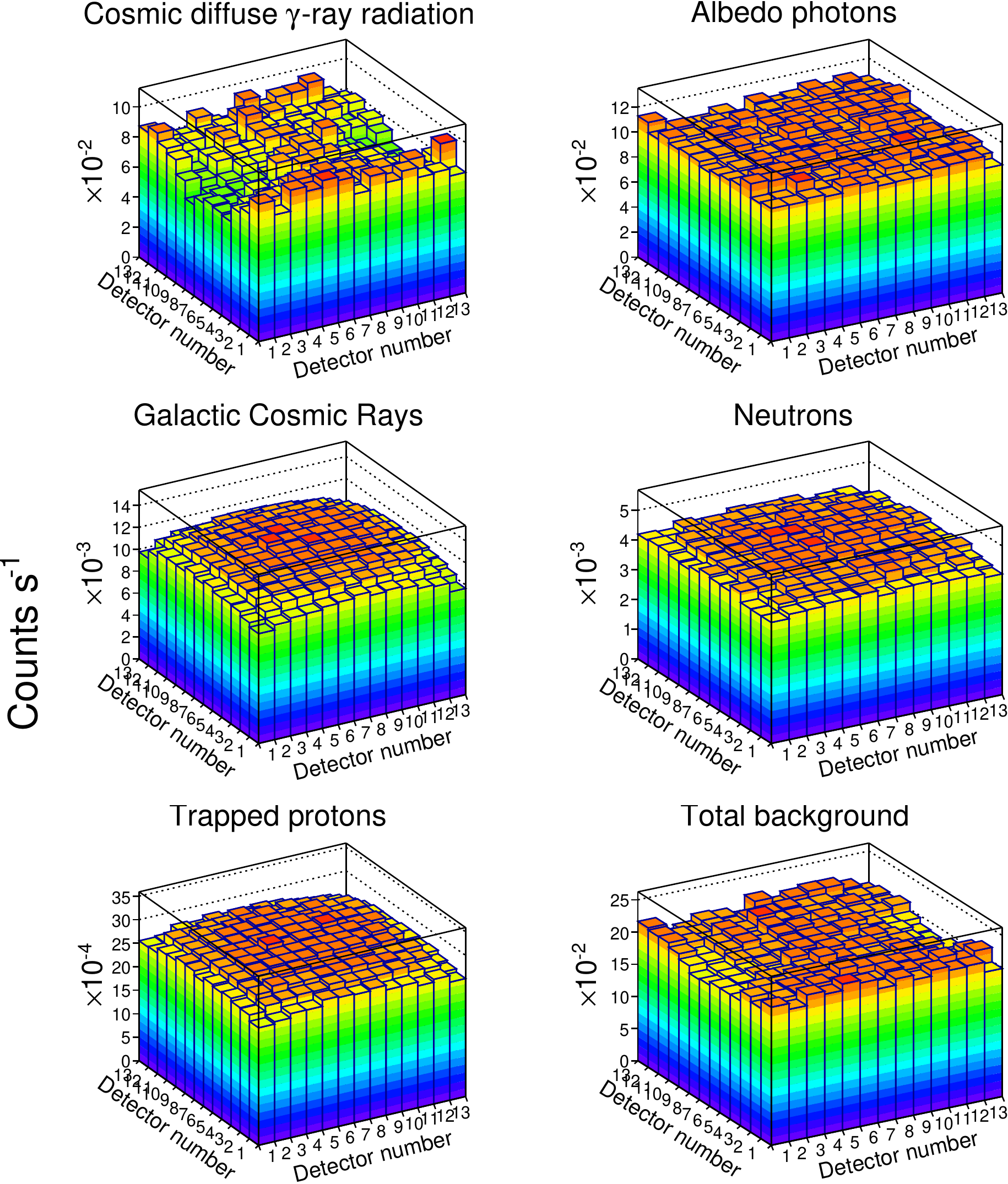}
\caption{Count rate distributions for each background component over the detection plane for {\it MIRAX\/}. The energy
range is 10\,-\,200\,keV. The typical 1-$\sigma$ error bars, in counts s$^{-1}$, are: diffuse $\gamma$ radiation:
4.7$\times 10^{-3}$; albedo: 3.0$\times 10^{-3}$; GCR: 1.2$\times 10^{-4}$; neutrons: 8.9$\times 10^{-5}$; trapped protons:
3.7$\times 10^{-5}$; total background: 5.6$\times 10^{-3}$.}
\label{fig:MIRAXcountdistr}
\end{figure*}

For {\it MIRAX\/}, the distributions (Figure \ref{fig:MIRAXcountdistr}) show similar structures. In this case, the 
predominance of the diffuse $\gamma$ radiation up to 40\ts keV makes the total background distribution less 
homogeneous. The count map due to albedo photons coming from below, which dominates above 40\ts keV, is uniform within 
statistics. In this case the count rate is 35.3 counts s$^{-1}$ in the total energy range 10\,-\,200\,keV. The similar 
result with respect to {\it protoMIRAX\/} is a coincidence due to the fact that the background is lower but the energy 
band is wider. In the 30--200\ts keV range, the same band of {\it protoMIRAX\/}, the total count rate for {\it MIRAX\/} 
is 22.4 counts\ts s$^{-1}$.

\begin{figure*}
\centering
\includegraphics[height=0.95\vsize]{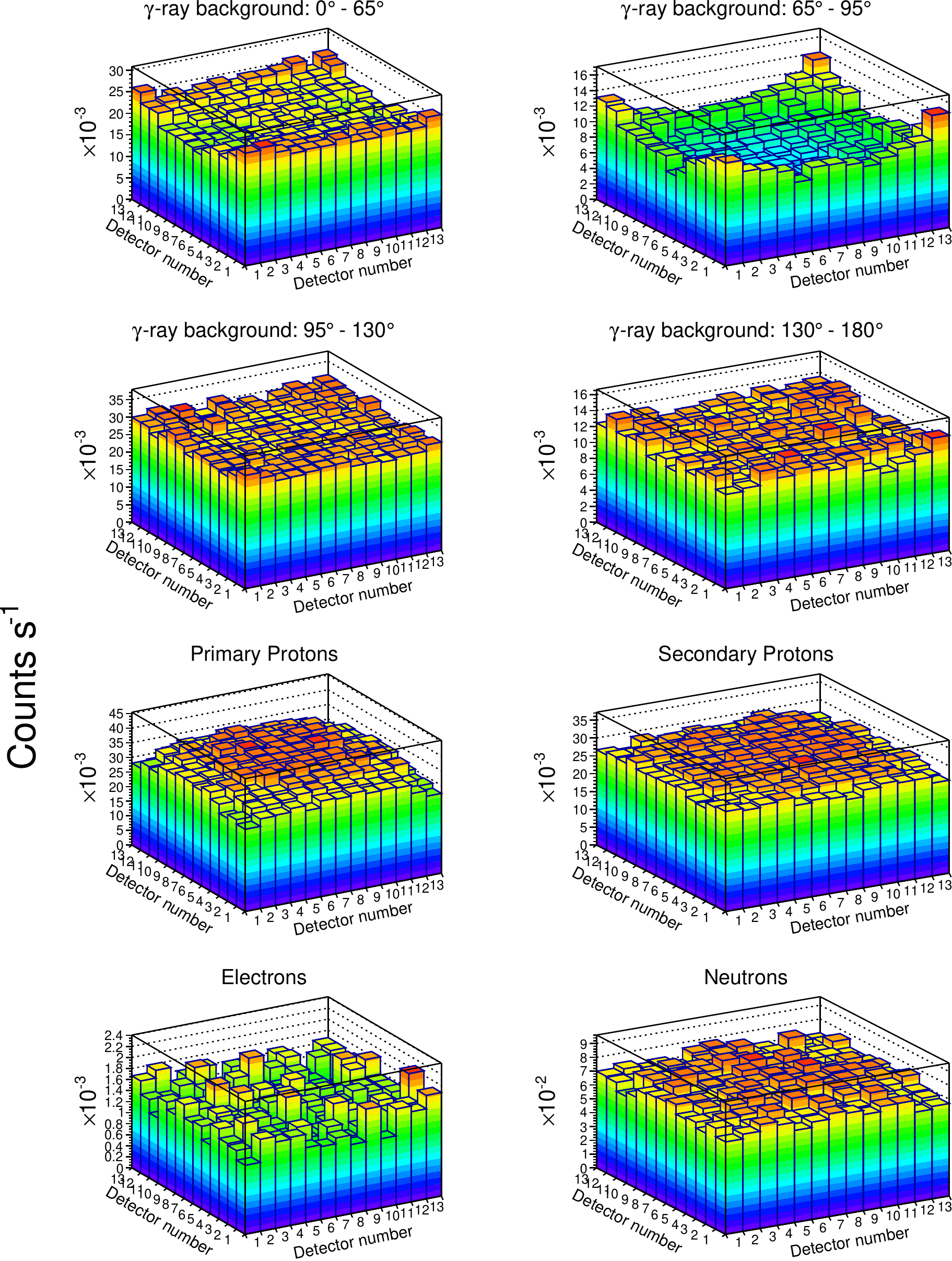}
\caption{Count rate distribution for each background component, over the full detection plane for {\it protoMIRAX}.
The simulated integration time is 8 hours for all distributions, and the energy range is 30--200\ts keV. The typical
1-$\sigma$ error bars (left-right, top-bottom) for the detectors are 8.8$\times 10^{-4}$,  5.2$\times 10^{-4}$, 1.0$\times 10^{-3}$,
6.7$\times 10^{-4}$, 1.0$\times 10^{-3}$, 1.0$\times 10^{-3}$, 2.2$\times 10^{-4}$, and 1.6$\times 10^{-3}$counts s$^{-1}$.}
\label{fig:shadowgram background protoMIRAX}
\end{figure*}

\begin{figure}
\centering
\includegraphics[width=0.9\hsize]{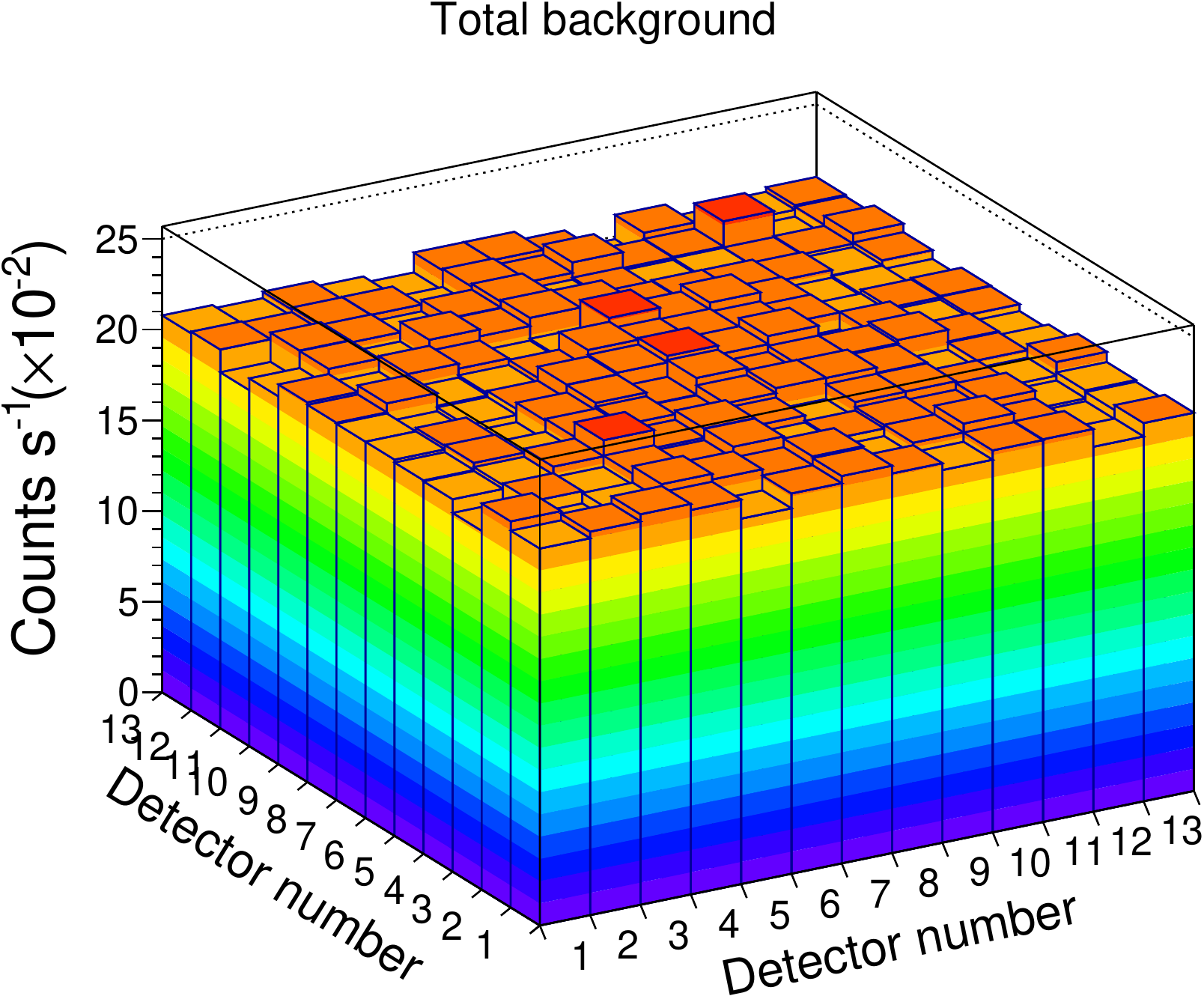}
\caption{Count rate distribution of the total background over the detection plane in the energy range 30\,-\,200\,keV.
The integration time is 8 hours at balloon altitudes. The typical 1-$\sigma$ error bar for the detectors is
2.8$\times 10^{-3}$ counts s$^{-1}$.}
\label{fig:total background shadowgram}
\end{figure}

\section{imaging reconstruction}
\label{sec:imaging reconstruction}

In order to study the effects of the background in the imaging reconstruction process, we have considered the presence 
in the field of view of point sources than can be observed by  {\it protoMIRAX\/} and {\it MIRAX\/}. The sources are 
the Crab nebula and two well-known Galactic black holes,  {\ONE} and {\GRS}, near the Galactic centre. The Crab's 
spectrum can be represented by a broken power-law model with $\Gamma_{1}=2.105$, $E_{\rm break}=100\,\text{keV}$, 
$\Gamma_{2}=2.22$ and a flux at 1 keV of 10.2 photons cm$^{-2}$ s$^{-1}$ keV$^{-1}$ \citep{Jourdain2008}.

\begin{figure}
 \includegraphics[width=\hsize]{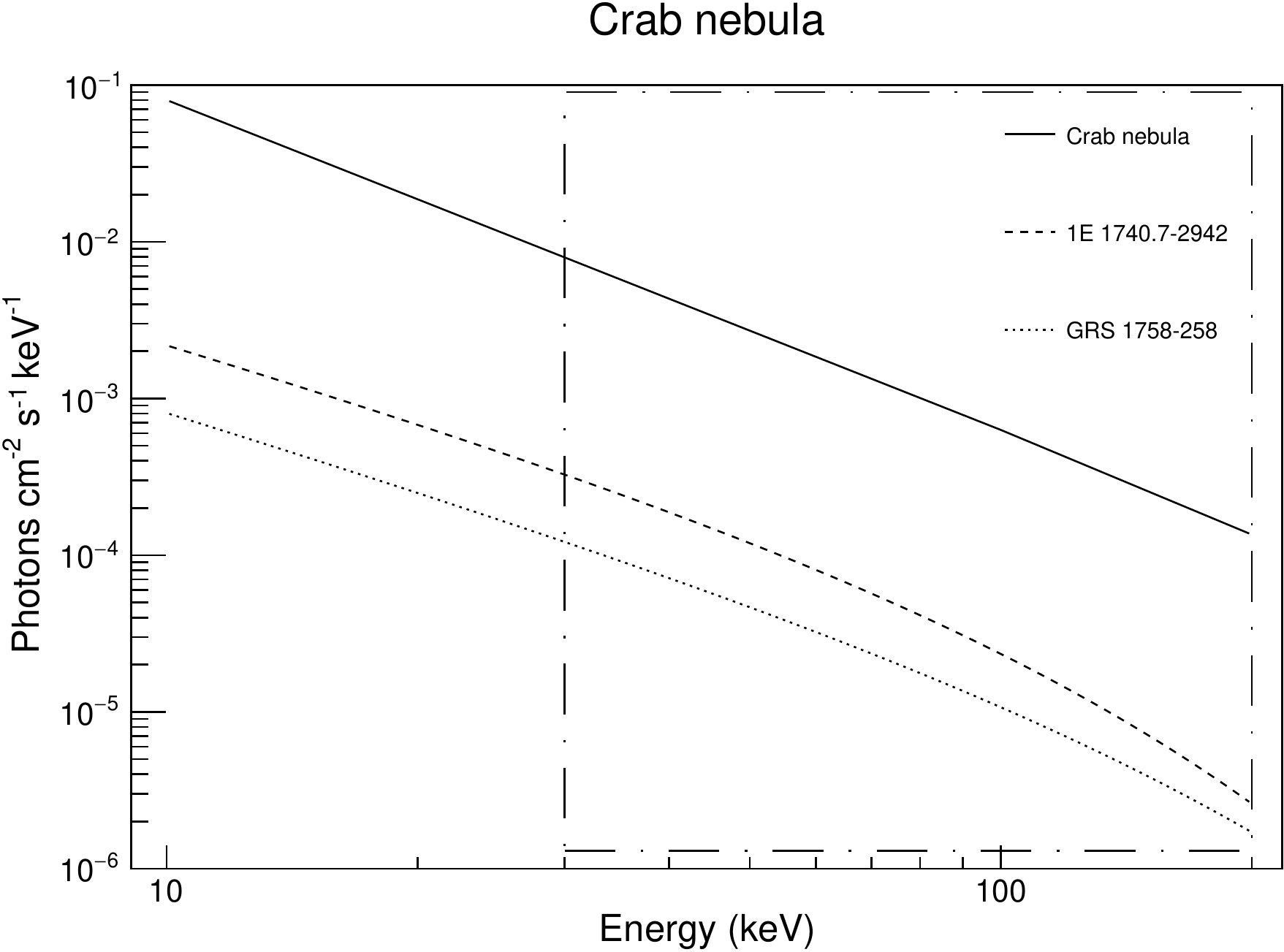}
\caption{Spectra used to simulate the emission from Crab, {\ONE} and {\GRS}. The dashed box represents the energy range
used to simulate the imaging reconstruction for {\it protoMIRAX\/} (considering attenuation by the atmosphere). For the
{\it MIRAX\/} imaging reconstruction, we have considered the full energy range (10\,-\,200\,keV).}
\label{ThreeSpectra}
\end{figure}

As {\ONE} and {\GRS} spend most of their time in the Low/Hard state \citep{Remillard2006}, their emission in 
X/$\gamma$-rays can be represented as a combination of a powerlaw and a disk emission (if necessary). The model 
parameters for each source are shown in Table \ref{sourcespectrum}. The three spectra are shown in Figure 
\ref{ThreeSpectra}.

\begin{table}
\caption{Spectra used for simulating the emission of {\ONE} and {\GRS}. The values are compatible with the sources in
the Low/Hard state.}
\label{sourcespectrum}
\begin{threeparttable}[b]
 \begin{center}
  \begin{tabular}{cccc}
\hline \hline
\multicolumn{4}{c}{\texttt{diskbb + cutoffpl}}  \\
\hline
					&				&	\ONE \tnote{a}		& 	\GRS \tnote{b}		\\ \hline
 \texttt{diskbb}			&	$T_{in}(\text{keV})$	&	0.15			&	0.67		\\
\hline
\multirow{3}{*}{\texttt{cutoffpl}}	&	$\Gamma$		&	1.52			&	1.59		\\
					&	$E_{\rm cut}(\text{keV})$	&	87.1			&	136		\\
\hline \hline
  \end{tabular}
 \begin{tablenotes}
  \item[a] \cite{Castro2014}.
  \item[b] \cite{Pottschmidt2006}.
 \end{tablenotes}
 \end{center}
\end{threeparttable}
\end{table}

\subsection{Imaging reconstruction for {\it protoMIRAX\/}}

\begin{figure*}
 \begin{tabular}{cc}
 \includegraphics[width=0.5\hsize]{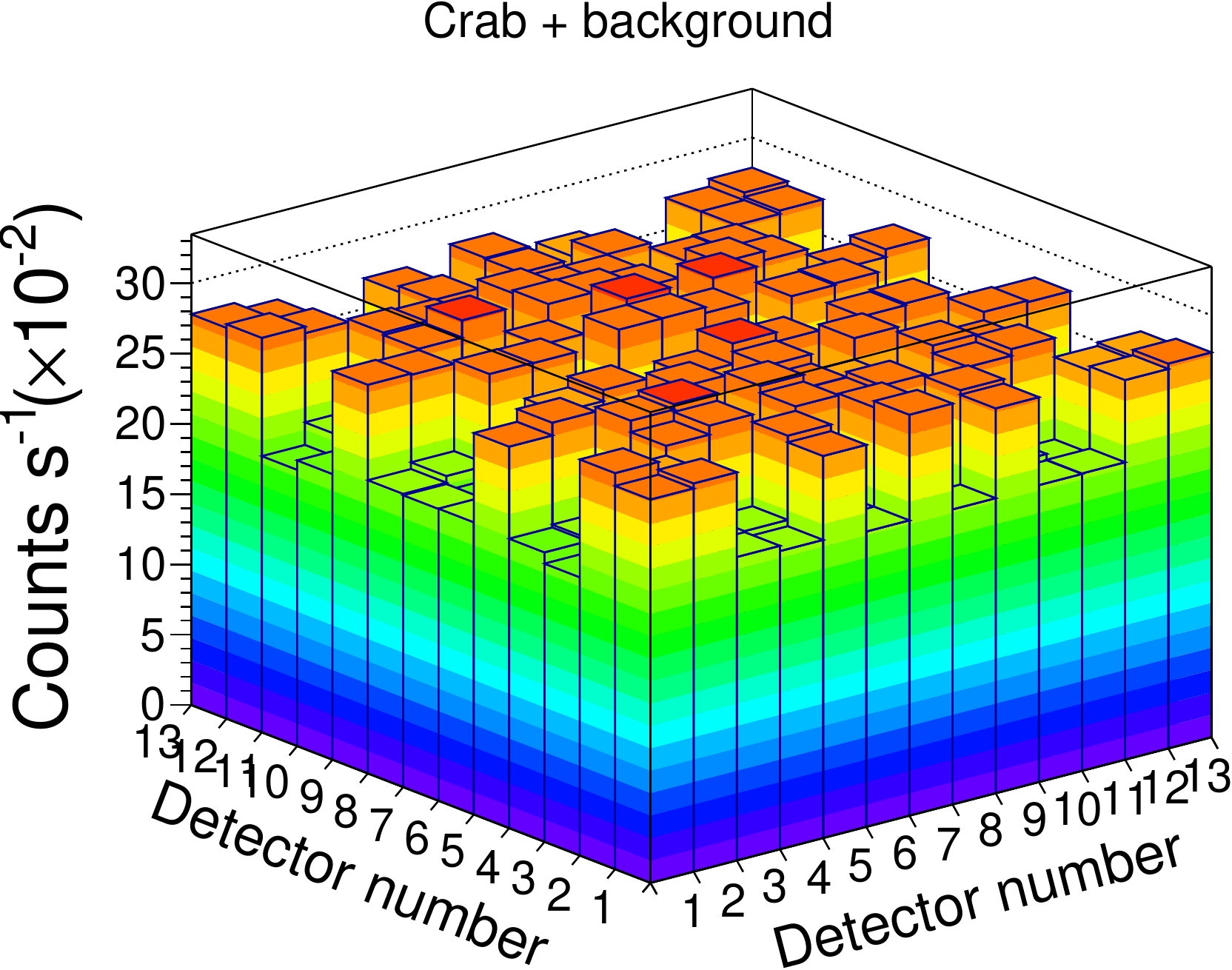}	& \includegraphics[width=0.5\hsize]{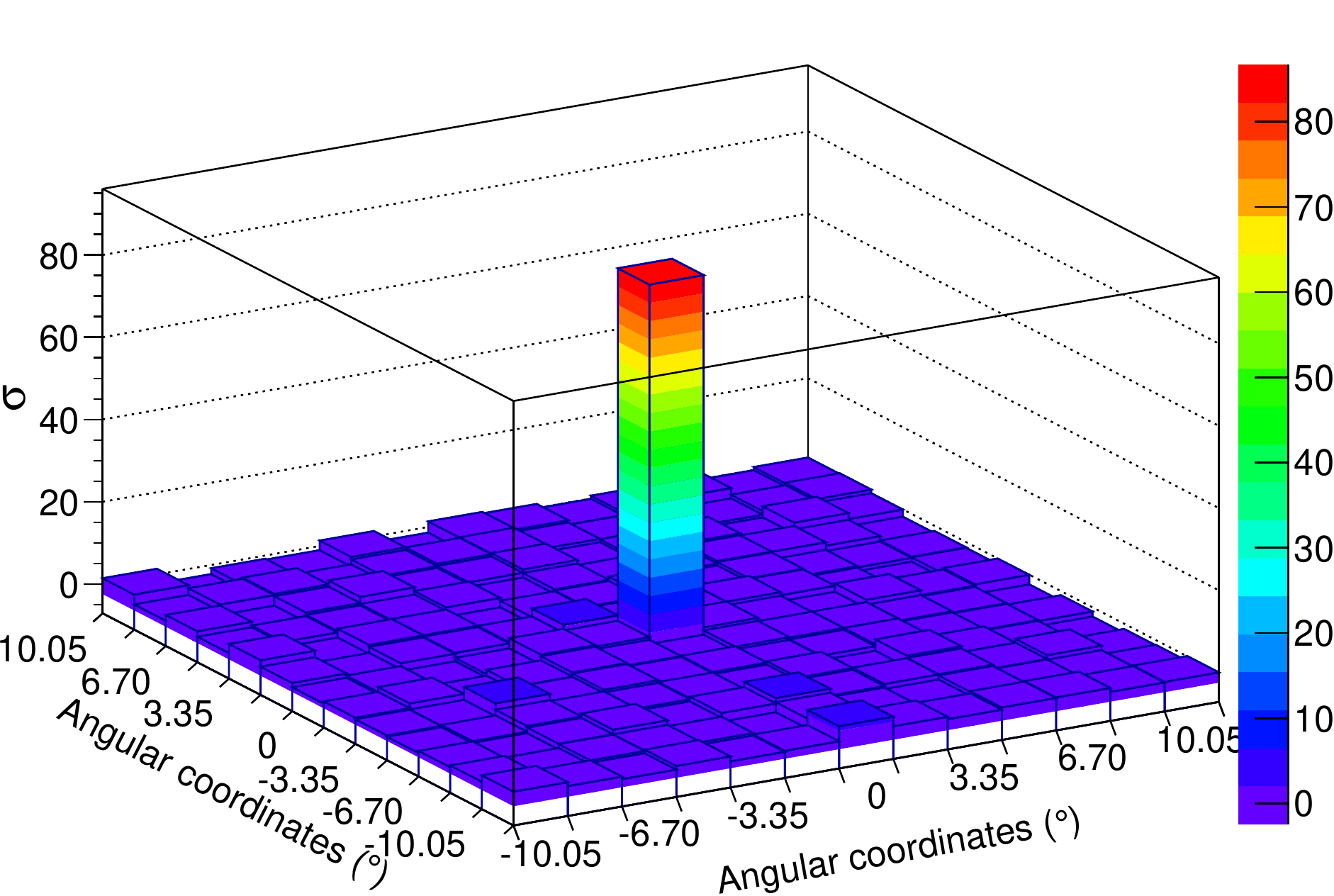}  \\
 \end{tabular}
\caption{Crab simulated image as observed by protoMIRAX during 4 hours around a meridian passage at a latitude of $-$23\gr. {\it Left:}  count map over the detector plane, including background. {\it Right:} reconstructed image in units of $\sigma$.}
\label{Reconst_Crab_proto_4h}
\end{figure*}

Since the {\it protoMIRAX\/} experiment will fly in the atmosphere at an altitude of $\sim$42\,km, the fluxes from the 
astrophysical sources will suffer atmospheric absorption according to:
\begin{equation}
 F(E)=F_{0}(E)\;e^{-\frac{\mu}{\rho}(E)\;x\;\text{sec}(z)}
\label{absorption}
\end{equation}
where $F_{0}$ is the flux at the top of the atmosphere, $\mu/\rho$ is the absorption coefficient of the 
atmosphere\footnote{\url{http://physics.nist.gov/PhysRefData/XrayMassCoef/ComTab/air.html}} in units of  cm$^{2}$ 
g$^{-1}$, $z$ is the zenith angle and $x$ is the expected atmospheric depth (2.7 g cm$^{-2}$) at an altitude of 
$\sim$42\,km. The zenith angle is a function of the geographical latitude $\varphi$ where the balloon will fly, the 
right ascension $\alpha$ and declination $\delta$ of the observed source, and the local sidereal time $t$. The zenith 
angle is given by:
\begin{equation}
 \cos z(t)=\sin\varphi\sin\delta + \cos\varphi\cos\delta\cos(t-\alpha)
\end{equation}
In this way, it is possible to compute the number of photons $N$ that will hit the detector coming from a specific 
point source, attenuated by the atmosphere. The number $N$ of photons passing through an area $S$, in a time interval 
$T$ and in the energy range $E_{\rm min}$ -- $E_{\rm max}$ is given by:
\begin{equation}
 N=\int_{S}\int_{E_{\rm min}}^{E_{\rm max}}\int_{0}^{T}F_{0}(E)\; e^{-\frac{\mu}{\rho}(E)\;x\;\text{sec}(z(t))} dt\,dE\,dS
\label{photonssource}
\end{equation}
We have simulated the Crab radiation by integrating Equation \ref{photonssource} in the energy range 30\,-\,200\,keV 
for a meridian passage of the source at a latitude of $-$23\gr. We considered a 4-hour interval centred at the 
meridian crossing. The Crab's reconstructed image is shown in Figure \ref{Reconst_Crab_proto_4h}. These high 
signal-to-noise ratio ($\sim80$) images of the Crab will be used to calibrate the imaging system and demonstrate the 
performance of the camera in a near-space environment. 

\subsection{Imaging reconstruction for {\it MIRAX\/}}

\begin{figure*}
 \begin{tabular}{cc}
  \includegraphics[width=0.5\hsize]{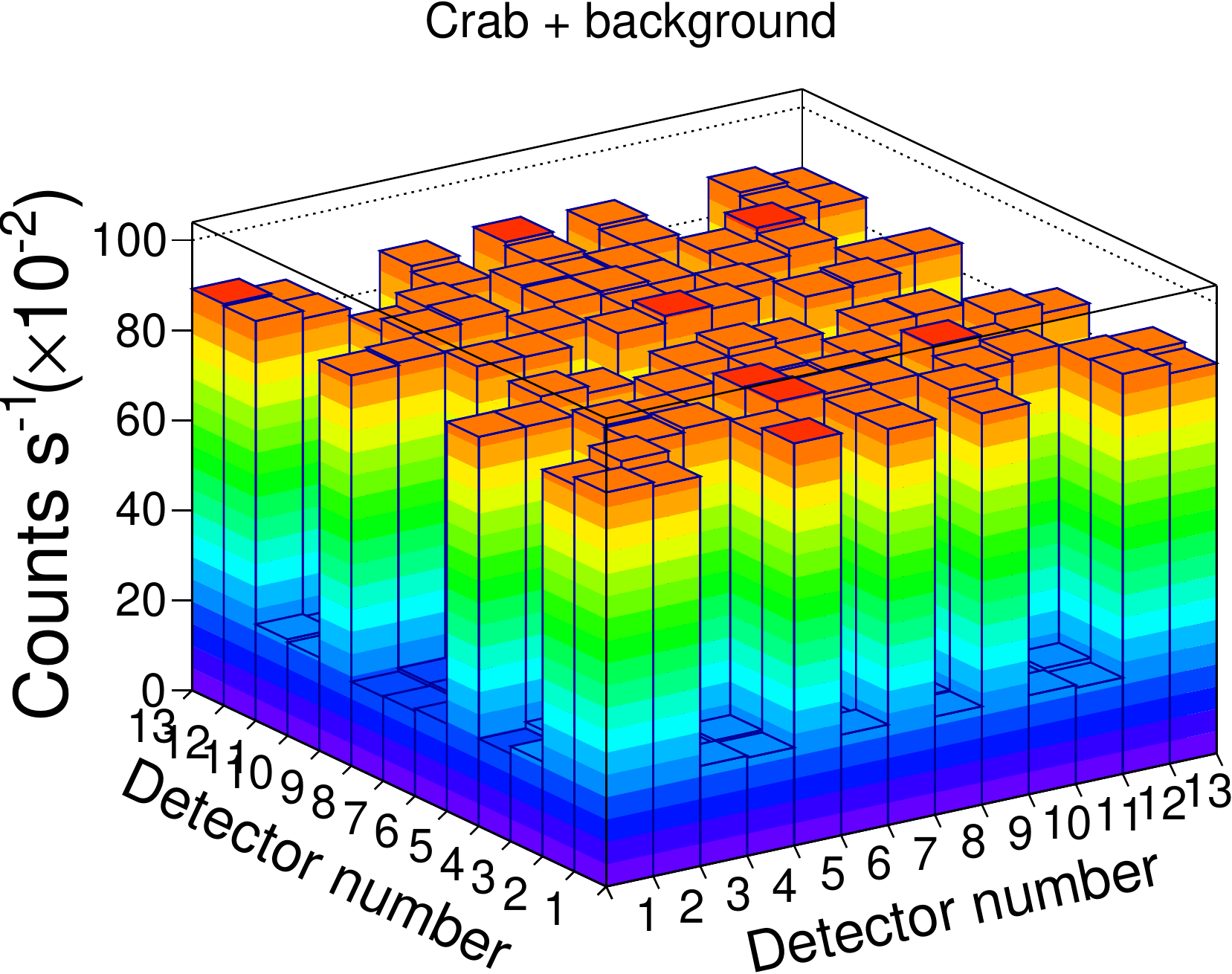} & \includegraphics[width=0.5\hsize]{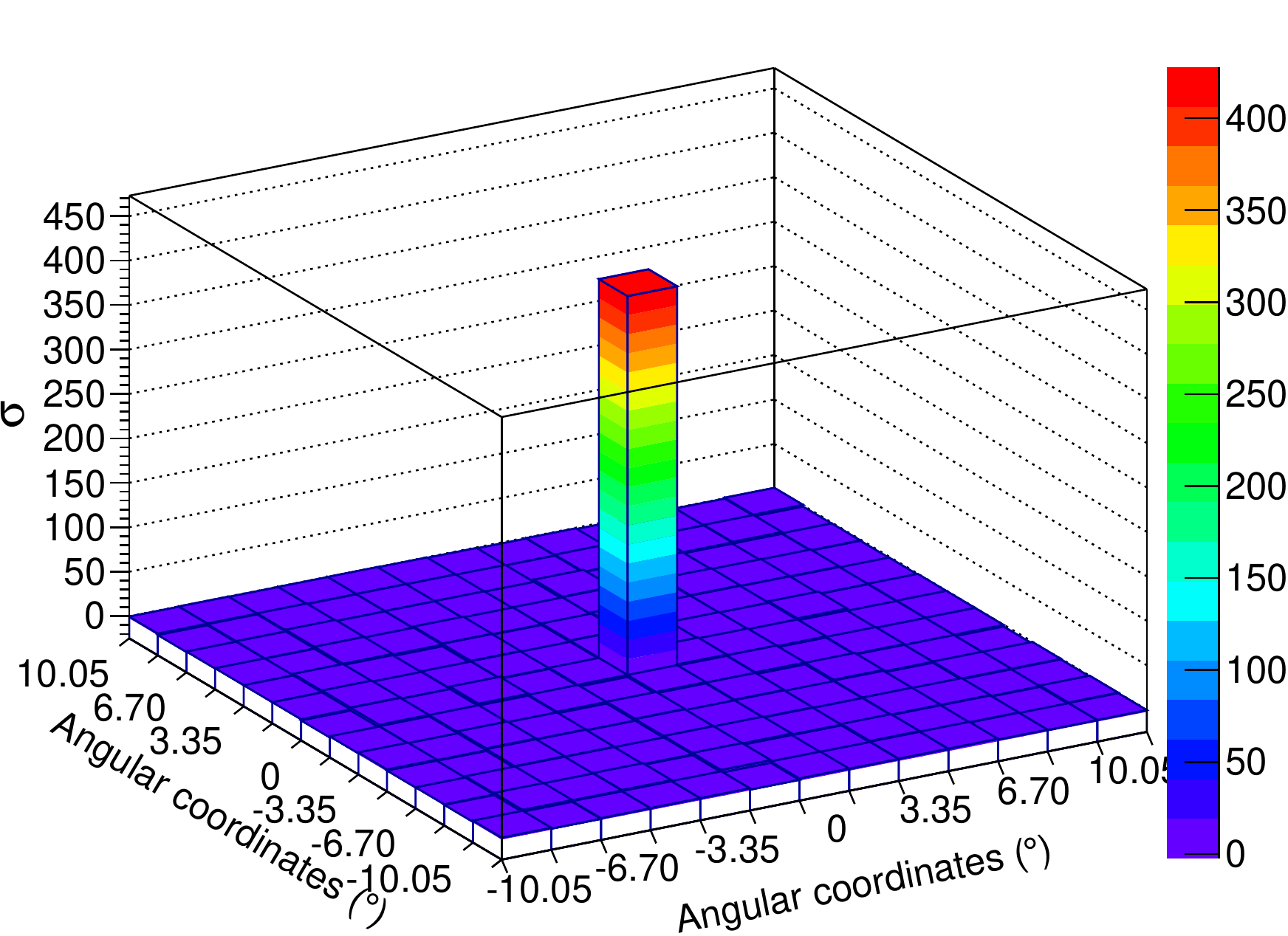}
 \end{tabular}
\caption{Imaging reconstruction of Crab as observed by {\it MIRAX\/}. {\it Left panel:} distribution of counts over the
detection plane considering both source and background. {\it Right panel:} reconstructed image for the Crab during a 4-h
observation.}
\label{CrabMIRAX}
\end{figure*}

\begin{figure}
 \includegraphics[width=\hsize]{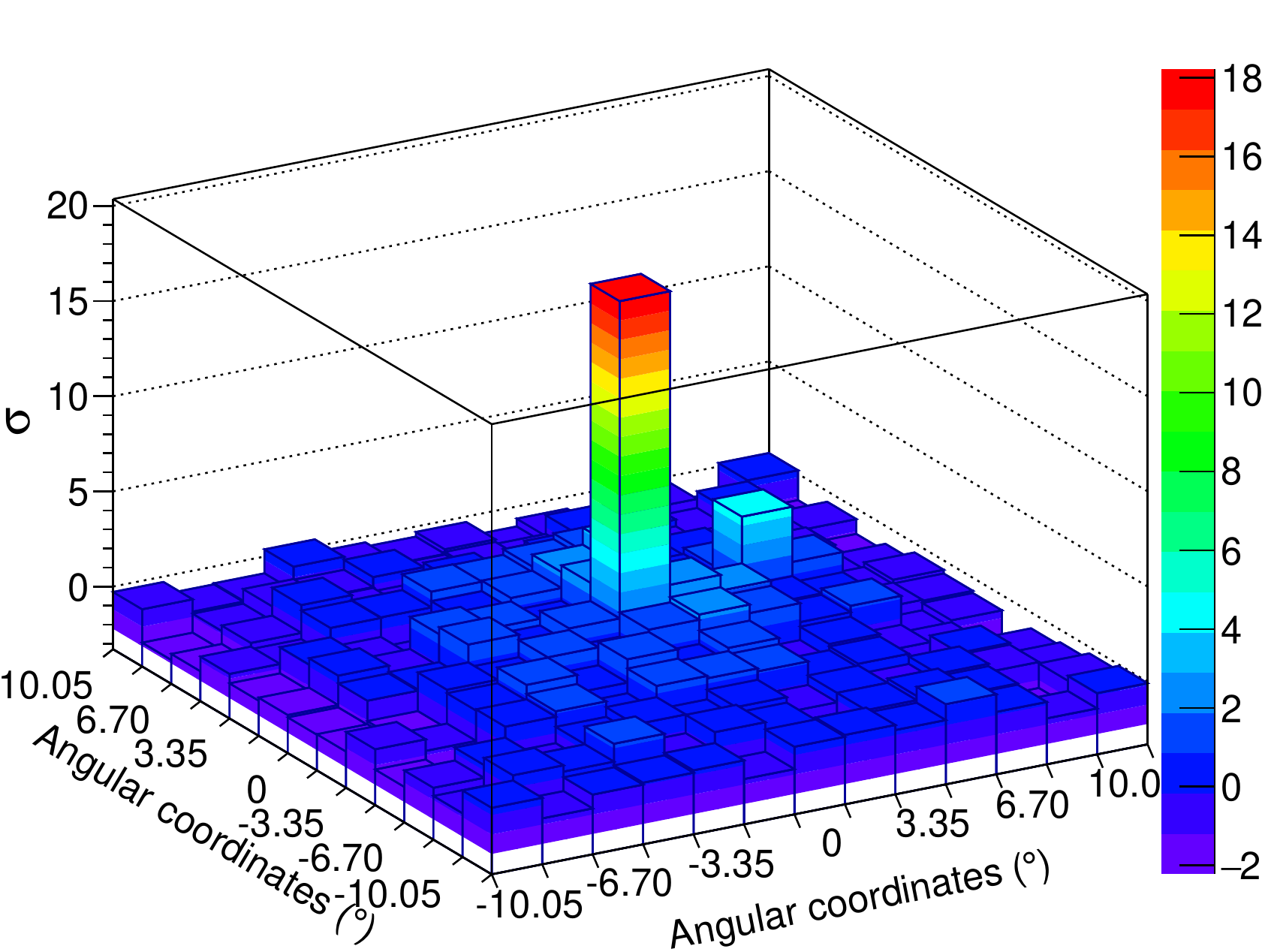}
\caption{GC reconstructed image with {\ONE} (in the centre) and  {\GRS} in the field, for an observation of 8 hours.}
\label{1E_GRS_MIRAX}
\end{figure}

In the case of {\it MIRAX\/}, we have considered the three sources described previously, but in a space environment (no 
atmospheric absorption). The reconstructed images are shown in Figures \ref{CrabMIRAX} (for the Crab) and 
\ref{1E_GRS_MIRAX} (for {\ONE} and {\GRS}). We can see a great improvement in the Crab image (4 hours of integration in 
both cases), due to the lower background in space and mainly due to the fact that the Crab is observed without 
atmospheric absorption by {\it MIRAX\/}. Nevertheless, it is noteworthy that the spatial non-uniformity of the background across the detector plane, mostly due to the fact that the cosmic diffuse gamma-ray component is not evenly distributed (see first panel of Fig.\ \ref{fig:MIRAXcountdistr}), produces a $\sim$40\% loss in the signal-to-noise. This was calculated by comparing our Crab image with another in which the total background is replaced by a flat count distribution (with statistical fluctuations) of the same level. The SNR that we obtain in the latter is 709, whereas in the Crab image of Fig.\ \ref{CrabMIRAX} the SNR is 428.

In the GC image, we have put \ONE\ in the centre of the FoV. The detected flux of \GRS\  is lowered by the collimator 
response. It is important to note that the purpose of these images is to carry out a preliminary demonstration of the 
imaging system for the X-ray camera in space, based on the simulations reported here. During the development of MIRAX, 
we are pursuing alternatives in terms of detector technology that will eventually allow us to have a detector plane 
with much larger area and higher spatial resolution. This will certainly greatly improve the performance of the camera 
in terms of sensitivity and angular resolution, and make it a competitive instrument capable of achieving the 
project's scientific objectives described by \cite{2006AIPC..840....3B}.

\section{Sensitivity} 

In \cite{Braga2015} we have shown a sensitivity curve for {\it protoMIRAX}. Here we present an improved version of it based on the new background spectral distribution
that we report in this work (section \ref{bkg_spectra}). In particular, the refined mass model for the experiment, including different thicknesses for shielding materials, 
allowed us to decrease significantly the overall background and the intensity of the lead fluorescence lines. This has lowered the minimum observable flux, especially from
70 to 90\ts keV, improving the sensitivity. In this work we have also produced a sensitivity curve for the space version of the camera ({\it MIRAX\/} mission). The sensitivity values are calculated as the minimum detectable fluxes (at a given statistical significance) for energy bins centred at each energy $E$:\begin{equation}
F_{\rm min} = \frac{2N_{\sigma}}{\epsilon(E)} \; \sqrt{\frac{B(E)}{A_{\rm det}\, T\, \Delta E}} \;\; {\rm photons\; cm^{-2} s^{-1} keV^{-1}},
\end{equation}
where $N_{\sigma}$ is the statistical significance in $\sigma$ units (or signal-to-noise ratio, in this case), $\epsilon(E)$ is the detector efficiency at energy $E$ (see \cite{Braga2015}), $B(E)$ is the background level in counts cm$^{-2}$ s$^{-1}$ keV$^{-1}$ at energy $E$, $A_{\rm det}$ is the geometrical area of the detector plane in cm$^2$, $T$ is the integration time in seconds, and $\Delta E$ is the energy band in keV. The factor $2$ in the above expression appears due to the fact that, for a coded aperture telescope with an open mask fraction of $0.5$, $A_{\rm eff} = \epsilon\, A_{\rm det}/2$, where $\epsilon$ is the detector efficiency.

\begin{figure*}
 \begin{tabular}{cc}
\includegraphics[width=0.5\hsize]{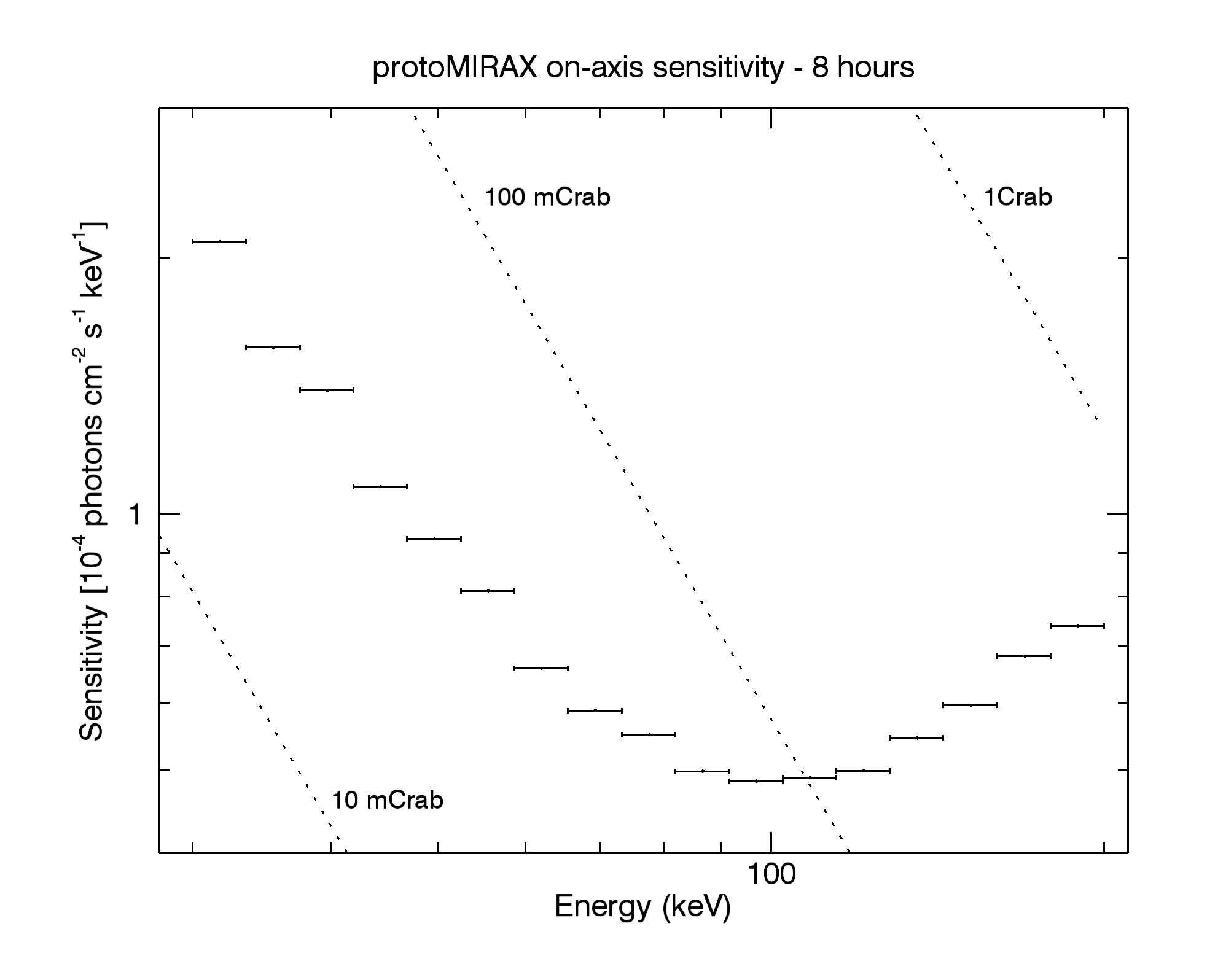}	& \includegraphics[width=0.5\hsize]{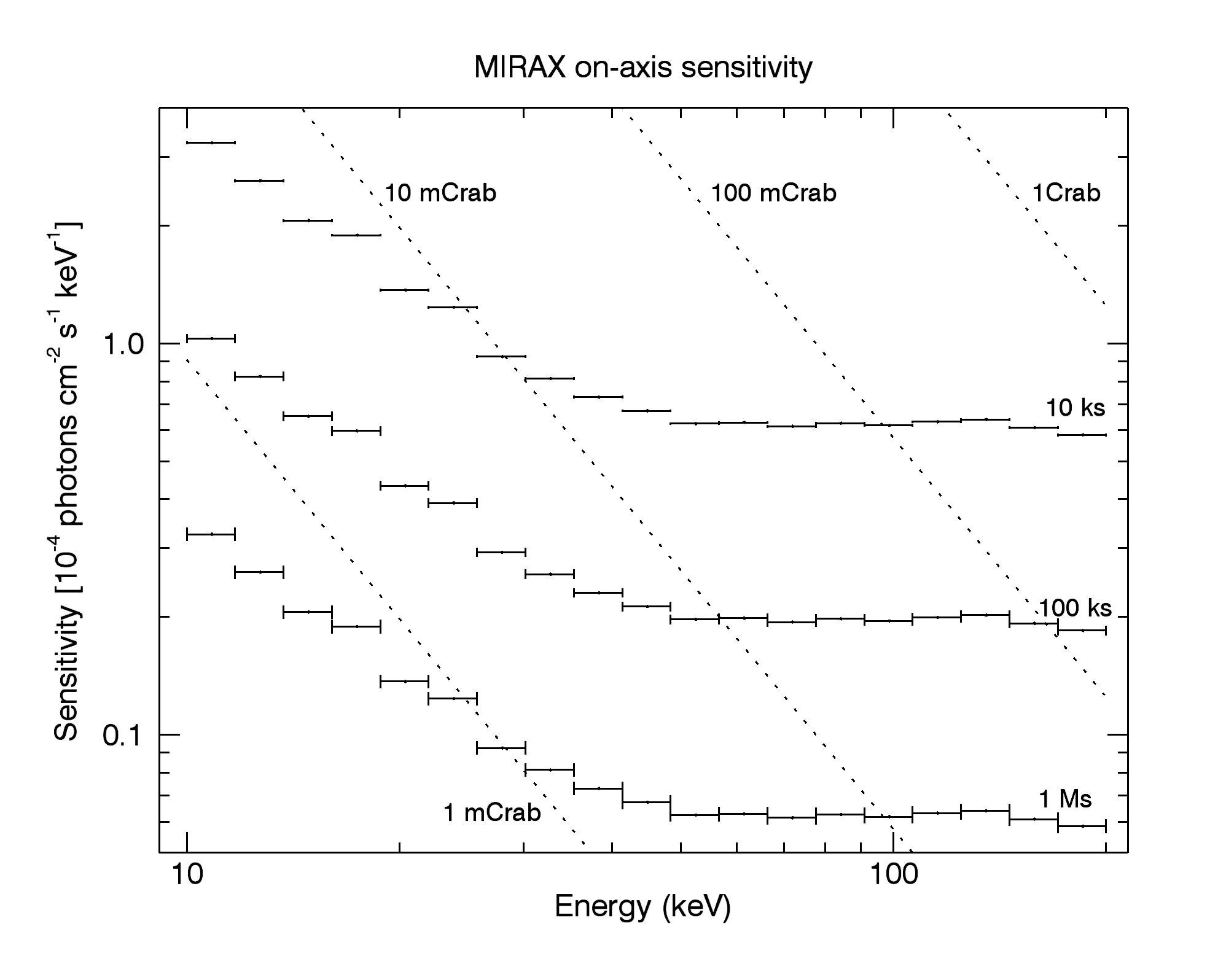} \\
\end{tabular}
\caption{ {\it Left Panel}: On-axis sensitivity curve for {\it protoMIRAX\/} for an integration of 8 hours at an atmospheric depth of 2.7 g cm$^{-2}$ and a latitude of $-$23\gr\ over Brazil. {\it Right Panel}: On-axis sensitivity curve for {\it MIRAX\/} for three different integration times at near-equatorial LEO. In both cases, the horizontal bars represent the minimum detectable flux at a statistical significance of 3\ts $\sigma$.} 
\label{sensitivity}
\end{figure*}

The sensitivity plots are shown in Figure \ref{sensitivity}.

\section{Discussion} \label{sec:discussion}

The simulations shown in this paper have provided very important informations and constraints for the optimal design of the hard X-ray camera we are planning to use as part of the {\it MIRAX\/} space program at INPE. The configuration we are currently developing, with a collimator and a coded mask, has shown to yield interesting results essentially due to two important reasons: (a) by using a collimator, we can make the full FoV of the instrument to be coded by the mask, avoiding the well-known undesirable effects of the ubiquitous partially coded fields of views in coded aperture imaging; and (b) the mounting of the camera become simpler and lighter since we avoid having to include large shielding flat panels on the sides of camera, joining the detector plane with the mask; a simple slim structure made of a light material can support the mask. In our case, the weight of the collimator is about half the weight of the shielding panels we would have to build for providing roughly the same shielding effectiveness. If we build cameras with larger distances between masks and detector to achieve better angular resolution, this becomes increasingly more important. However, the use of a collimator also brings undesirable properties since the instrument's sensitivity will decrease towards the edges of the FoV according to the collimator response. In our case, the angular FWHM response provided by the collimator is 14$^{\circ}$ and imaging beyond that will be severely limited. Another problem is that the collimator produces a distribution of counts in the detector plane that is spatially non-uniform and depend on the incident directions of the components that are responsible for the detector background. We have shown here that this is effect is more significant when the camera is placed in the low orbit environment due to the intense cosmic X-ray diffuse radiation (CDR).
 
The optimum design of a coded aperture imager will depend on the scientific objectives of the experiment. If a survey instrument is preferred, the use of collimators will probably not be the best option due to the limitations on the sensitivity over the FoV. If, on the other hand, the objective is to obtain sharp images of selected fields with no interference from adjacent fields with bright sources, a collimator could be the best option.

The simulations presented here have shown that, on hard X-ray balloon experiments, the background induced by neutrons is very important below 50 keV. Above this energy, the background produced by gamma rays and protons are equally important, and electrons make up a negligible contribution. At LEO, the CDR is by far the most important contribution to the background up to $\sim$40 keV. Above that, the contribution form albedo radiation becomes dominant. The analysis of these relative contributions is very important to the design of shielding structures and materials for X-ray experiments.
 
\section{Conclusions} \label{sec:conclusion}

The design and development of space astronomy X- and $\gamma$-ray experiments require a reliable estimation of the 
background levels against which the sources of interest will be observed. Particularly, in coded aperture experiments a 
large detector area is fundamental for achieving competitive sensitivities. In those cases the backgrounds are usually 
very intense and show inhomogeneities across the detector plane due to geometrical factors.

In this paper we show detailed Monte Carlo simulations of the background and imaging observations of a hard X-ray 
imaging camera that is being developed in the scope of the {\it MIRAX\/} space mission. This instrument is a prototype 
that is going to be tested in stratospheric balloon flights. The {\it MIRAX\/} mission will play a very important role 
in the study of hard X-ray sources and transient phenomena, since it will constantly monitor a large area around the 
central Galactic plane region.

We present separate spectra and count distribution across the detector plane for each relevant radiation field 
impinging in the instrument and discuss their contribution. We also take into account the angular distribution of the 
different incident photons and particles.  We provide detailed background characteristics and levels for two different 
environments in which the camera is supposed to operate: stratospheric balloon altitudes and near-equatorial low-Earth 
orbit. 

The simulations reported here allowed us to define with high accuracy the best configuration for the shielding 
structure and coded mask materials of the instrument. In order to minimize the intense lead fluorescence lines in the 
range of $\sim$70-80\ts keV that we have seen in the initial simulations, we were able to obtain an optimised 
configuration of materials for both the shielding and the mask. Also, the initial concentration of counts towards the 
edges and corners of the detector plane was completely overcome by a different geometry of both the external collimator 
blades and the placement of side shielding materials. The simulations have provided us with the precise count rates 
expected when we operate the instrument at balloon altitudes and LEO, and enabled the calculation of detailed 
sensitivities as a function of  energy. 

To the best of our knowledge, this is the first time a detailed description of GEANT4 simulations for space astronomy 
instruments is reported. We hope that this will be of great value for researchers and instrument developers in this 
field.

\section*{Acknowledgements}

We thank FINEP for financial support under Conv\^enio 01.10.0233.00. We also thank CNPq and FAPESP, Brazil, for support
 under INCT Estudos do Espa\c{c}o. A.P. acknowledges the support by the international Cooperation Program CAPES-ICRANET
financed by CAPES - Brazilian Federal Agency for Support and Evaluation of Graduate Education, within the Ministry of
Education of Brazil.  M.C. acknowledges the support of INPE/CNPq through the PCI program  and R.S. acknowledges the support by CAPES, Brazil.




\bibliography{refs}
\bibliographystyle{mnras}



\appendix
%
%


\bsp	
\label{lastpage}
\end{document}